\pdfoutput=1

\documentclass[sigconf,screen]{acmart}
\setcopyright{acmlicensed}
\acmPrice{15.00}
\acmDOI{10.1145/3611643.3616337}
\acmYear{2023}
\copyrightyear{2023}
\acmSubmissionID{fse23main-p1004-p}
\acmISBN{979-8-4007-0327-0/23/12}
\acmConference[ESEC/FSE '23]{Proceedings of the 31st ACM Joint European Software Engineering Conference and Symposium on the Foundations of Software Engineering}{December 3--9, 2023}{San Francisco, CA, USA}
\acmBooktitle{Proceedings of the 31st ACM Joint European Software Engineering Conference and Symposium on the Foundations of Software Engineering (ESEC/FSE '23), December 3--9, 2023, San Francisco, CA, USA}
\received{2023-02-02}
\received[accepted]{2023-07-27}

\usepackage{tikz}
\usepackage{pgfplots}
\pgfplotsset{width=9cm,compat=1.9}

\usepackage{amsmath}
\usepackage{caption}
\usepackage{subcaption}
\usepackage{amsfonts}
\usepackage{array}
\usepackage{wrapfig}
\usepackage{calculator}
\usepackage{xspace}
\usepackage{multirow}
\usepackage{enumitem}
\usepackage[altpo]{backnaur}
\usepackage[ruled, vlined, linesnumbered, commentsnumbered, longend]{algorithm2e}
\usepackage{pifont}%
\newcommand{\cmark}{\ding{51}}%
\newcommand{\xmark}{\ding{55}}%
\usepackage{fp}
\newcommand\ppercent[2]{\FPeval\result{round(#1*100/#2,1)}\result\%}

\setlist[itemize]{leftmargin=*}
\setlist[enumerate]{leftmargin=*}

\SetCommentSty{mycommfont}

\usetikzlibrary{calc,fit,shapes}
\newcommand*\myc[1]{%
\scalebox{0.78}{\begin{tikzpicture}[baseline=-4pt]
  \node[draw,circle,inner sep=0.5pt, fill=black] {\textcolor{white}{\textsf{\textbf{#1}}}};
\end{tikzpicture}}}

\usepackage[normalem]{ulem}
\definecolor{applegreen}{rgb}{0.55, 0.71, 0.0}

\usepackage{libertine}
\usepackage{graphicx}
\usepackage{framed}
\usepackage[subtle]{savetrees}

\newcommand*\openquote{\makebox(15,-20){\scalebox{5}{``}}}
\newcommand*\closequote{\makebox(15,-20){\scalebox{5}{''}}}
\definecolor{azure}{rgb}{0.94, 1.0, 1.0}
\colorlet{shadecolor}{azure}
\newenvironment{myquote}[1]%
  {\list{}{\leftmargin=#1\rightmargin=#1}\item[]}%
  {\endlist}

\makeatletter
\newif\if@right
\def\shadequote{\@righttrue\shadequote@i}
\def\shadequote@i{\begin{snugshade}\begin{myquote}{1cm}\openquote}
\def\endshadequote{%
  \if@right\hfill\fi\closequote\end{myquote}\end{snugshade}}
\@namedef{shadequote*}{\@rightfalse\shadequote@i}
\@namedef{endshadequote*}{\endshadequote}
\makeatother

\begin{document}

\date{}

\title{
\sys: Diversifying DNN Generation via Inductive Rule Inference
}
\settopmatter{authorsperrow=4}
\author{Jiawei Liu}
    \affiliation{\institution{University of Illinois}\city{Urbana-Champaign}\country{USA}}
    \email{jiawei6@illinois.edu}
\author{Jinjun Peng}\authornote{The work was performed during a remote internship at University of Illinois.}
    \affiliation{\institution{Columbia University}\city{New York}\country{USA}}
    \email{jinjun.peng@columbia.edu}
\author{Yuyao Wang}\authornotemark[1]
    \affiliation{\institution{Nanjing University}\city{Nanjing}\country{China}}
    \email{yuyao6@outlook.com}
\author{Lingming Zhang}
    \affiliation{\institution{University of Illinois}\city{Urbana-Champaign}\country{USA}}
    \email{lingming@illinois.edu}

\newcommand{\gir}{GraphIR\xspace}
\newcommand{\sys}{\textsc{NeuRI}\xspace}
\newcommand{\sysr}{\textsc{NeuRI$^r$}\xspace}
\newcommand{\sysi}{\textsc{NeuRI$^i$}\xspace}
\newcommand{\FreeFuzz}{FreeFuzz\xspace}
\newcommand{\nnsmith}{\textsc{NNSmith}\xspace}

\newcommand{\freefuzz}{FreeFuzz\xspace}
\newcommand{\doctor}{DocTer\xspace}
\newcommand{\deeprel}{DeepREL\xspace}
\newcommand{\cradle}{Cradle\xspace}
\newcommand{\lemon}{LEMON\xspace}
\newcommand{\audee}{Audee\xspace}

\newcommand{\muffin}{Muffin\xspace}
\newcommand{\Rosette}{Rosette\xspace}

\newcommand{\eg}{\emph{e.g.,}\xspace}
\newcommand{\ie}{\emph{i.e.,}\xspace}

\newcommand{\subst}[2]{[{#1}/{#2}]\xspace}
\newcommand{\insta}[1]{{#1}^\star\!\xspace}

\newcommand{\RuleInst}{Partial operator\xspace}
\newcommand{\RuleInsts}{Partial operators\xspace}
\newcommand{\ruleInst}{partial operator\xspace}
\newcommand{\ruleInsts}{partial operators\xspace}

\newcommand{\op}{\phi\xspace}
\newcommand{\attr}{A\xspace}
\newcommand{\opt}{T\xspace} %
\newcommand{\gram}{\mathcal G\xspace}
\newcommand{\eset}{\mathcal E\xspace}
\newcommand{\Records}{\mathcal R\xspace}
\newcommand{\opi}{{\hat\phi}\xspace}
\newcommand{\shape}{d\space}
\newcommand{\Shape}{D\space}
\newcommand{\icons}{\mathcal C \xspace}
\newcommand{\sprop}{\mathcal P \xspace}
\newcommand{\dinp}{I \xspace}
\newcommand{\dout}{O \xspace}
\newcommand{\ico}{c \xspace}
\newcommand{\spr}{p \xspace}
\newcommand{\expr}{\text{\sf expr}\xspace}
\newcommand{\rinp}{\mathcal I \xspace}

\newcommand{\Dim}{d\space}
\newcommand{\Dims}{D\space}

\newcommand{\OUTER}{\textcolor{orange}{\texttt{LABEL}}}

\newcommand{\ePT}{PyTorch\xspace}
\newcommand{\eTF}{TensorFlow\xspace}
\newcommand{\CovTFImproveOp}{7\%\xspace}
\newcommand{\CovPTImproveOp}{2\%\xspace}
\newcommand{\CovTFImproveModel}{24\%\xspace}
\newcommand{\CovPTImproveModel}{15\%\xspace}
\newcommand{\CovTFPerc}{10.8\%\xspace}
\newcommand{\CovPTPerc}{17.4\%\xspace}

\newcommand{\tfops}{1370\xspace}
\newcommand{\xlaops}{450\xspace}
\newcommand{\jitops}{1310\xspace}

\newcommand{\locFuzzer}{11.6k\xspace}
\newcommand{\locRmFuzzer}{7.9k\xspace}
\newcommand{\locFuzzerCore}{1.1k\xspace}
\newcommand{\locSynthesizer}{1.5k\xspace}
\newcommand{\locInstrumenter}{1.8k\xspace}

\newcommand{\locTotal}{14.9k\xspace}
\newcommand{\locTotalNet}{7k\xspace}

\newcommand{\PTBugFixIcons}{17}  %
\newcommand{\PTBugFixCE}{25}    %
\newcommand{\PTBugFixSan}{7}    %
\newcommand{\PTBugFixByNonSan}{2}    %
    \ADD{\PTBugFixIcons}{\PTBugFixSan}\PTBugFixIS
\ADD{\PTBugFixIS}{\PTBugFixCE}\PTBugFixTemp %
\ADD{\PTBugFixTemp}{\PTBugFixByNonSan}\PTBugFix %

\ADD{\PTBugFixIcons}{2}\PTBugConfirmIcons   %
\ADD{\PTBugFixCE}{7}\PTBugConfirmCE        %
\ADD{\PTBugFixSan}{6}\PTBugConfirmSan       %
\ADD{\PTBugFixByNonSan}{1}\PTBugConfirmByNonSan       %
    \ADD{\PTBugConfirmIcons}{\PTBugConfirmSan}\PTBugConfirmIS
\ADD{\PTBugConfirmIS}{\PTBugConfirmCE}\PTBugConfirmTemp
\ADD{\PTBugConfirmTemp}{\PTBugConfirmByNonSan}\PTBugConfirm

\newcommand{\PTBugTriageIcons}{0}
\newcommand{\PTBugTriageCE}{9}
\newcommand{\PTBugTriageSan}{7}
    \ADD{\PTBugTriageIcons}{\PTBugTriageSan}\PTBugTriageIS
\ADD{\PTBugTriageIS}{\PTBugTriageCE}\PTBugTriage

\newcommand{\PTBugWontIcons}{0}
\newcommand{\PTBugWontCE}{2}
\newcommand{\PTBugWontSan}{0}
    \ADD{\PTBugWontIcons}{\PTBugWontSan}\PTBugWontIS
\ADD{\PTBugWontIS}{\PTBugWontCE}\PTBugWont

\ADD{\PTBugConfirm}{\PTBugWont}\PTBugCW
\ADD{\PTBugTriage}{\PTBugCW}\PTBug

    \ADD{\PTBugConfirmIcons}{\PTBugTriageIcons}\PTTempIcons
\ADD{\PTTempIcons}{\PTBugWontIcons}\PTBugIcons
    \ADD{\PTBugConfirmCE}{\PTBugTriageCE}\PTTempCE
\ADD{\PTTempCE}{\PTBugWontCE}\PTBugCE
    \ADD{\PTBugConfirmSan}{\PTBugTriageSan}\PTTempSan
\ADD{\PTTempSan}{\PTBugWontSan}\PTBugSan

\newcommand{\TFBugFixIcons}{0}  %
\newcommand{\TFBugFixCE}{0}    %
\newcommand{\TFBugFixSan}{0}    %
    \ADD{\TFBugFixIcons}{\TFBugFixSan}\TFBugFixIS
\ADD{\TFBugFixIS}{\TFBugFixCE}\TFBugFix %

\ADD{\TFBugFixIcons}{6}\TFBugConfirmIcons   %
\ADD{\TFBugFixCE}{7}\TFBugConfirmCE        %
\ADD{\TFBugFixSan}{1}\TFBugConfirmSan       %
    \ADD{\TFBugConfirmIcons}{\TFBugConfirmSan}\TFBugConfirmIS
\ADD{\TFBugConfirmIS}{\TFBugConfirmCE}\TFBugConfirm

\newcommand{\TFBugTriageIcons}{0}
\newcommand{\TFBugTriageCE}{0}
\newcommand{\TFBugTriageSan}{0}
    \ADD{\TFBugTriageIcons}{\TFBugTriageSan}\TFBugTriageIS
\ADD{\TFBugTriageIS}{\TFBugTriageCE}\TFBugTriage

\newcommand{\TFBugWontIcons}{1}
\newcommand{\TFBugWontCE}{0}
\newcommand{\TFBugWontSan}{0}
    \ADD{\TFBugWontIcons}{\TFBugWontSan}\TFBugWontIS
\ADD{\TFBugWontIS}{\TFBugWontCE}\TFBugWont

\ADD{\TFBugConfirm}{\TFBugWont}\TFBugCW
\ADD{\TFBugTriage}{\TFBugCW}\TFBug

    \ADD{\TFBugConfirmIcons}{\TFBugTriageIcons}\TFTempIcons
\ADD{\TFTempIcons}{\TFBugWontIcons}\TFBugIcons
    \ADD{\TFBugConfirmCE}{\TFBugTriageCE}\TFTempCE
\ADD{\TFTempCE}{\TFBugWontCE}\TFBugCE
    \ADD{\TFBugConfirmSan}{\TFBugTriageSan}\TFTempSan
\ADD{\TFTempSan}{\TFBugWontSan}\TFBugSan

\ADD{\PTBug}{\TFBug}\ALLBug
\ADD{\PTBugFix}{\TFBugFix}\ALLBugFix
\ADD{\PTBugConfirm}{\TFBugConfirm}\ALLBugConfirm
\ADD{\PTBugTriage}{\TFBugTriage}\ALLBugTriage
\ADD{\PTBugWont}{\TFBugWont}\ALLBugWont

\ADD{\PTBugSan}{\TFBugSan}\ALLBugSan
\ADD{\ALLBugSan}{3}\ALLBugBypro %
\SUBTRACT{\ALLBug}{\ALLBugBypro}\ALLBugFuzzing
    \ADD{\PTBugIcons}{\PTBugCE}\PTBugFuzzing

\begin{abstract}
Deep Learning (DL) is prevalently used in various industries to improve decision-making and automate processes, driven by the ever-evolving DL libraries and compilers. The correctness of DL systems is crucial for trust in DL applications.
As such, the recent wave of research has been studying the automated synthesis of test-cases (\ie DNN models and their inputs) for fuzzing DL systems.
However, existing model generators only subsume a limited number of operators, for lacking the ability to pervasively model operator constraints.
To address this challenge, we propose \sys, a fully automated approach for generating valid and diverse DL models composed of hundreds of types of operators.
\sys adopts a three-step process:
(i) collecting valid and invalid API traces from various sources;
(ii) applying inductive program synthesis over the traces to infer the constraints for constructing valid models;
and
(iii) using hybrid model generation which incorporates both symbolic and concrete operators.
Our evaluation shows that \sys improves branch coverage of TensorFlow and PyTorch by \CovTFImproveModel and \CovPTImproveModel over the state-of-the-art model-level fuzzers.
\sys finds \ALLBug{} \emph{new} bugs for PyTorch and TensorFlow in four months, with \ALLBugConfirm{} already fixed or confirmed.
Of these, 9 bugs are labelled as \emph{high priority} or \emph{security vulnerability}, constituting 10\% of all high-priority bugs of the period.
Open-source developers regard error-inducing tests reported by us as ``high-quality'' and ``common in practice''.
\end{abstract}



\date{}
\maketitle

\section{Introduction}

The rise of Deep-Learning (DL) libraries and compilers has enabled emerging AI applications, such as AI chatbots~\cite{chatgpt}, art generators~\cite{dalle} and autonomous driving, powering hundreds of millions of users.
These complex systems have become increasingly adopted and ever evolving.
For example, PyTorch~\cite{pytorch} and TensorFlow~\cite{tensorflow}, the most popular DL systems with 62k and 171k GitHub stars respectively,
are moving toward their next major version (\ie PyTorch 2~\cite{pt2} and TensorFlow 3~\cite{tf3}),
aiming at better model compilation support.
However, taking PyTorch's new compiler~\cite{dynamo} as an example, since birth (\ie 17 months) it is insufficiently tested by a test suite in \textit{eight} thousand LoC. 
Consequently, it is crucial to harness the correctness of DL systems via extensive and automated testing.

\noindent\textbf{The test-case generation problem} for DL systems is to synthesize a DNN model and its computational inputs.
Additionally, generating \textit{diverse} and \textit{valid} models is essential for making high-quality tests.%

\begin{enumerate}
    \item \emph{Model diversity}: 
        Effective DL system testing asks for model diversity coming from the variety of APIs, as well as the way they are composed.
        Additionally, to test the complicated DL compilers, it is important to generate models with multiple operators of various types for practicing the compiler passes~\cite{nnsmith}.
    \item \emph{Validity}:
        DNN models are programs~\cite{tf-security-program} -- for well-formedness they need to comply with validity constraints.
        Arbitrarily constructing and composing operators,
        such as creating pooling operators with negative kernel sizes or ``connecting'' operators with unwanted tensor shapes,
        oftentimes violate the constraints for constructing a well-defined model.
        As a result, argument errors (for DL libraries) or parser errors (for DL compilers) are raised before deeper system behaviours are tested.
\end{enumerate}

\newcommand{\Tsapiclr}{red!50!white}
\newcommand*\Tsapi{%
\scalebox{0.5}{\begin{tikzpicture}[baseline=-4pt]
  \node[draw,regular polygon,regular polygon sides=3,fill=\Tsapiclr] {};
\end{tikzpicture}}}

\newcommand{\Tswocclr}{white!70!blue}
\newcommand*\Tswoc{%
\scalebox{0.75}{\begin{tikzpicture}[baseline=-4pt]
  \node[draw,circle,fill=\Tswocclr] {};
\end{tikzpicture}}}

\newcommand{\TnnsmithClr}{yellow!30!orange}
\newcommand*\TnnsmithShape{%
\scalebox{0.75}{\begin{tikzpicture}[baseline=-4pt]
  \node[draw,diamond,fill=\TnnsmithClr] {};
\end{tikzpicture}}}

\newcommand{\TGoalClr}{yellow!40!green}
\newcommand*\TGoalShape{%
\scalebox{0.75}{\begin{tikzpicture}[baseline=-4pt]
  \node[draw,regular polygon,regular polygon sides=5,fill=\TGoalClr] {};
\end{tikzpicture}}}

\setlength\intextsep{0pt}
\setlength{\columnsep}{6pt}
\setlength{\abovecaptionskip}{0pt}

\begin{wrapfigure}[15]{r}{0pt}
\centering
\begin{tikzpicture}
\begin{axis}[
xmin=1.15, xmax=6.15, ymin=0.3, ymax=5.4,
x=0.8cm, y=1cm,
ylabel shift = -5.5mm,
ylabel style={rotate=-180, xshift = -30pt},
xlabel style={xshift = 10pt, yshift = 13pt},
axis x line=bottom, axis y line=left,
xlabel = {\large \bf Operator Diversity}, ylabel = {\large \bf Model Diversity},
ticks=none,
mark size=5pt,
]
\addplot[domain=0:4.8][line width=4pt, dotted, color=yellow!50!green, opacity=.35]{x};

\addplot[mark=triangle*,style={solid, fill=\Tsapiclr}] coordinates {(4.8, 0.8)}; %
\node [left] at (axis cs: 4.7,0.8) {\sf Single-API}; %

\addplot[mark=diamond*,style={solid, fill=\TnnsmithClr}] coordinates {(2.5, 2.5)}; %
\node [right] at (axis cs: 2.54,2.5) {\sf Strongly constrained}; %

\addplot[mark=*,style={solid, fill=\Tswocclr}] coordinates {(1.85, 1.4)}; %
\node [right] at (axis cs: 2,1.4) {\sf Weakly constrained}; %

\node [right,color=\TnnsmithClr] (destination) at (axis cs: 2.2, 1.85){\sf Manual operator rules.};
\draw[->][line width=2pt,color=\TnnsmithClr](axis cs: 1.95, 1.65)--(axis cs: 2.35, 2.3);

\draw[->][line width=2pt,color=\TGoalClr,](axis cs: 2.75, 2.85)--(axis cs: 4.5, 4.6) 
node [fill=red!10!white,rounded corners,inner sep=2.2pt,semithick, %
midway,pos=.5,sloped,above=1.8mm,text width=2.72cm,text=red] {{\bf This paper:}\\ \sf Auto. rule inference!};

\addplot[mark=pentagon*,style={solid, fill=green!70!gray}] coordinates {(4.8, 4.8)}; %
\node[above] at (axis cs: 5.15, 5.05) {\large \bf Our Goal};

\end{axis}
\end{tikzpicture}
\caption{Test-case diversity.}\label{fig:diversity}
\end{wrapfigure}

\noindent\textbf{Motivation.}
The model diversity primarily depends on the comprehensiveness of operators, which are the building blocks to a model.
Prior work on single-API testing~\cite{wei2022free,xie2022docter} can generate a large body of API invocations (including both operator and utility APIs) via mutation or generation which comply with high-level type constraints or the plausible value sets. 
Can we directly apply such high-level information to generate valid DNN models?
Unfortunately, it is impractical.
Because constructing a valid API invocation further requires satisfying fine-grained constraints between operator attributes (\ie non-tensor arguments such as kernel sizes and strides) and input tensor types\footnote{Following prior work~\cite{nnsmith}, a tensor type is a tuple of its shape and data type.} (particularly shapes).
For example, while single-API testers may understand that \texttt{conv2d} accepts an image and a weight tensor of floating-points (\ie type constraints),
their \emph{newly} created \texttt{conv2d} invocations are not guaranteed to have the channel dimension of the image matching that of the weight (\ie shape constraints).
Consequently, such attributes violate the validity properties required by \texttt{conv2d} and lead to invocation failures.
Without understanding such fine-grained constraints, it is unlikely to correctly compose various APIs for constructing well-formed and diverse models.
Intuitively, in Figure~\ref{fig:diversity} prior single-API testers \Tsapi{} can achieve ideal API diversity when APIs being validly constructed.
However, the diversity hardly extends at model-wise which requires multiple APIs to be constructed and ``connected'' correctly simultaneously.

Meanwhile, there are two categories of proposals for constructing valid models.
\emph{Weakly constrained model generation} \Tswoc{}~\cite{lemon,luo2021graph,muffin} limits the use of APIs to those with simple and straight-forward constraints.
For example, LEMON~\cite{lemon} only uses shape-preserving operators that have no input constraints.
Consequently, such operators can be arbitrarily constructed and added to build a model.
More recent work~\cite{luo2021graph,muffin} additionally inserts ``reshaping'' layers such that reshaped output tensors can stay in compatible shapes.
However, it still may not construct operators with valid attributes (non-tensor arguments).
Even worse, 
using such ``layer wrappers'' compromises the structural diversity of the models, which can overlook compiler passes activated by specific patterns.
To support more diverse APIs correctly, \nnsmith~\cite{nnsmith}, as a \textit{strongly constrained} \TnnsmithShape{} approach, defines a specification for describing input constraints and shape propagation (elaborated in~\S\ref{sec:rule}). 
Nonetheless, it requires manual efforts for specifying those rules.
For example, while a DL framework, \eg PyTorch, can define over two thousand APIs,
only about sixty are supported by \nnsmith after its first-year development.
Hence, it can take years for \nnsmith~\cite{nnsmith} to completely support a framework, \ie from \TnnsmithShape{} to \TGoalShape{}, which is unscalable.

\noindent\textbf{Insight.}
Can we scale the diversity of model generation by enabling more operators (\eg by hundreds) \emph{fully automatically}?
We start to answer this question from two insights:
\emph{(i)} Empirically we observed that most operator rules are simple, \eg consisting of arithmetic expressions for shape computation and if-else branches for handling conditions incurred by some attributes.
As a result, it is feasible to search a program that functions as operator rules, given the size of the problem is acceptable.
Specifically,
by instrumenting DL API invocations,
we can obtain a set of input-output examples,
with which the inference of operator rules can be regarded as a inductive program synthesis problem~\cite{Winston:1970,Lau:1998}.
\emph{(ii)}
Can an operator still be used for model generation even if its operator rule is not available?
We find it \emph{feasible} by inserting a ``concrete'' operator initialized by recorded invocation traces.
To make use of both symbolically and concretely obtained operators, we can apply a \textit{concolic} model generation approach to construct models with both sources.

\noindent\textbf{Summary.} This work makes the following contributions:
\begin{itemize}
\item 
    In this work, we present the urgency for improving API diversity of model generation and formally introduce the essential properties for generating valid DNNs -- \emph{operator rules}.
    Furthermore, we open the first proposal of automatically inferring operator rules for diversifying and scaling valid model generation.
\item
    We build \sys (\underline{\textsc{Neu}}ral Network Synthesis via \underline{\textsc{R}}ule \underline{\textsc{I}}nference), a fuzzer for testing DL systems with three steps:
    (i) an instrumenter that collects and augments API invocations from various sources; 
    (ii) an optimized rule synthesizer that efficiently infers operator rules with inductive program synthesis;
    and
    (iii) a hybrid model generator that compiles both symbolic and concrete information for producing valid and diverse DNNs.
\item
    We extensively and rigorously evaluated \sys.
    Within four months, \sys finds \ALLBug{} \emph{new} bugs for PyTorch and TensorFlow, with \ALLBugConfirm{} fixed or confirmed.
    9 of the PyTorch bugs are labelled as \emph{high priority} or \emph{security vulnerability}, constituting around 10\% of all high-priority bugs in PyTorch's bug tracker of the period.
    By evaluating branch coverage, \sys improves the state-of-the-art model-level fuzzer by \CovPTImproveModel (\ePT{}) / \CovTFImproveModel (\eTF{}).
\end{itemize}

\section{Operator Rules}\label{sec:rule}

The functionality of a deep-learning model (\ie DNN) can be represented as a list of operations, each of which transforms one or multiple input tensors (\ie multi-dimensional arrays) to output tensors.
Accordingly, a test-case in DL systems constructs a DNN and is evaluated over some computational inputs, expecting the model can be successfully executed and produce correct results.

For generating effective test-cases automatically, it is crucial to generate and diversify valid DNN models.
State-of-the-art \nnsmith~\cite{nnsmith} constructs valid DNNs
with operator constraints and shape propagation rules.
With SMT solvers, such rules can help statically construct an operator which can be safely inserted to a given model.
Because in DL frameworks such operator rules are implicit defined and cannot be exported directly, they are manually specified in \nnsmith.
However, crafting them from scratch is unscalable.
For example, in the first-year development of \nnsmith, only around sixty operators are implemented with rules, despite the fact that many rules are even repetitive.
As a result, for diversifying operators being used and saving manual effort of domain experts, we aim at inferring those operator rules \emph{automatically}. 
We now formalize and elaborate the operator rules:

\newcommand{\opexamp}{\texttt{avg\_pool2d}\xspace}
\newcommand{\kernelh}{\texttt{kh}\xspace}
\newcommand{\kernelw}{\texttt{kw}\xspace}
\newcommand{\padh}{\texttt{padh}\xspace}

\begin{figure}
    \centering
    \includegraphics[width=\linewidth]{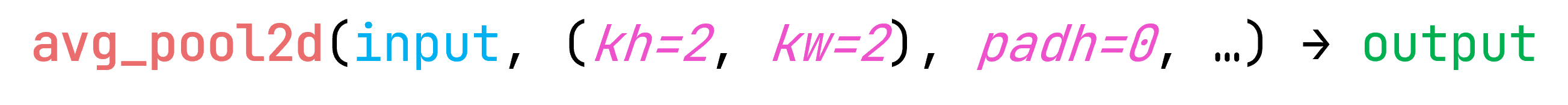}
    \caption{The symbolic view of \opexamp.}
    \label{fig:rexamp}
\end{figure}

\noindent\textbf{Symbolizing operators.}
As is shown in Figure~\ref{fig:rexamp}, an operator is a function which takes input tensors (\eg \texttt{input}) and configurations (\eg \texttt{kh}) as arguments.
The configurations, also known as operator attributes, describe high-level semantics for performing an operation and can impact the operator rules.
For example, \texttt{kw} and \texttt{kh} define the size for applying the ``avg'' filter over the input image which must be no smaller than the kernel size (assume no padding).
For being evaluated \textit{statically}, operator rules only leverage and symbolize an operator's compile-time information: (i) operator type (\eg \opexamp), (ii) $\dinp$ which is a list of input shape vectors, and (iii) $\attr$ as the set of operator attributes.
Runtime information such as the detailed element values inside the input tensors is too costly to be modelled. 
Meanwhile, we use $\dout$ to denote the output shapes produced by the shape propagation rule.

\noindent\textbf{Rule \#1: Input constraints.}
The input constraints of an operator are a set of \textit{predicate functions} $\icons = \{\ico_1, \ico_2, \cdots\}$ over $\attr \cup \dinp$. %
For example, constraints in \opexamp require the kernel size to be no larger than the padded image size (\eg $i_h + 2\times a_{\padh}$), namely:

$$
\ico_k\left(\attr=\{a_{\kernelh}, a_{\padh},\cdots\}, \dinp=\left[[i_c, i_h, i_w]\right]\right) = a_{\kernelh} \le i_h + 2a_{\padh}
$$

The arguments of $\ico_k$ consist of the attributes to \opexamp and the shape list with the shape of the only input tensor (\ie $|I|=1$).
It is worth noting that for clarity we assume the input is a non-batch image with only three dimensions (\ie the channel, height and width); however, in practice, 2d-pooling also accepts batched inputs with an extra batch dimension.
Meanwhile, some operators could take a variable length of inputs (\eg concatenate)
or outputs (\eg split).
For being general, a predicate may not assume the \emph{tensor signature}, \ie \# of input/output tensors and their ranks, to be fixed.
Thus, operator rules with such patterns may need to be described with conditional branches (\eg the syntax of PyTea~\cite{jhoo2022static}).
We later in \S\ref{sec:inf} introduce how to leverage \emph{\ruleInsts{}} to simplify such branches which are difficult to handle in rule inference.

\newcommand{\spropexamp}{\texttt{avg\_pool2d}\xspace}

\noindent\textbf{Rule \#2: Shape propagation.}
Because evaluating Rule 1 requires knowing the dimensions of operator inputs, which are outputs of other operators from the DNN under construction, output shapes of operators also need to be evaluated.
The shape propagation rule for an operator can be described by a function $\sprop$ over $\attr \cup \dinp$, which returns a list of propagated output shapes as $\dout$.
For instance, the shape propagation for \spropexamp can be described as:

\begin{align}
\begin{split}
&\sprop\left(\attr=\{a_{\kernelh}, a_{\padh},\cdots\}, \dinp=\left[[i_c, i_h, i_w]\right]\right) = [[o_c, o_h, o_w]] \\
&\text{where }
\begin{cases}
o_c =& i_c \\
o_h =& \left\lfloor\frac{i_h + 2\times a_{\texttt{padh}} - a_{\texttt{kh}}}{a_{\texttt{stride}}} + 1\right\rfloor \\
o_w =& \left\lfloor\frac{i_w + 2\times a_{\texttt{padw}} - a_{\texttt{kw}}}{a_{\texttt{stride}}} + 1\right\rfloor
\end{cases}
\end{split}
\label{eq:poolshape}
\end{align}

For example, given input shape of [3,3,3], we can tell the corresponding output shape for operator in Figure~\ref{fig:rexamp} (assuming $a_\texttt{stride}$ is 1) is [3,2,2] without invoking it.

\section{Approach}\label{sec:approach}

\begin{figure*}
    \centering
    \includegraphics[width=0.9\linewidth]{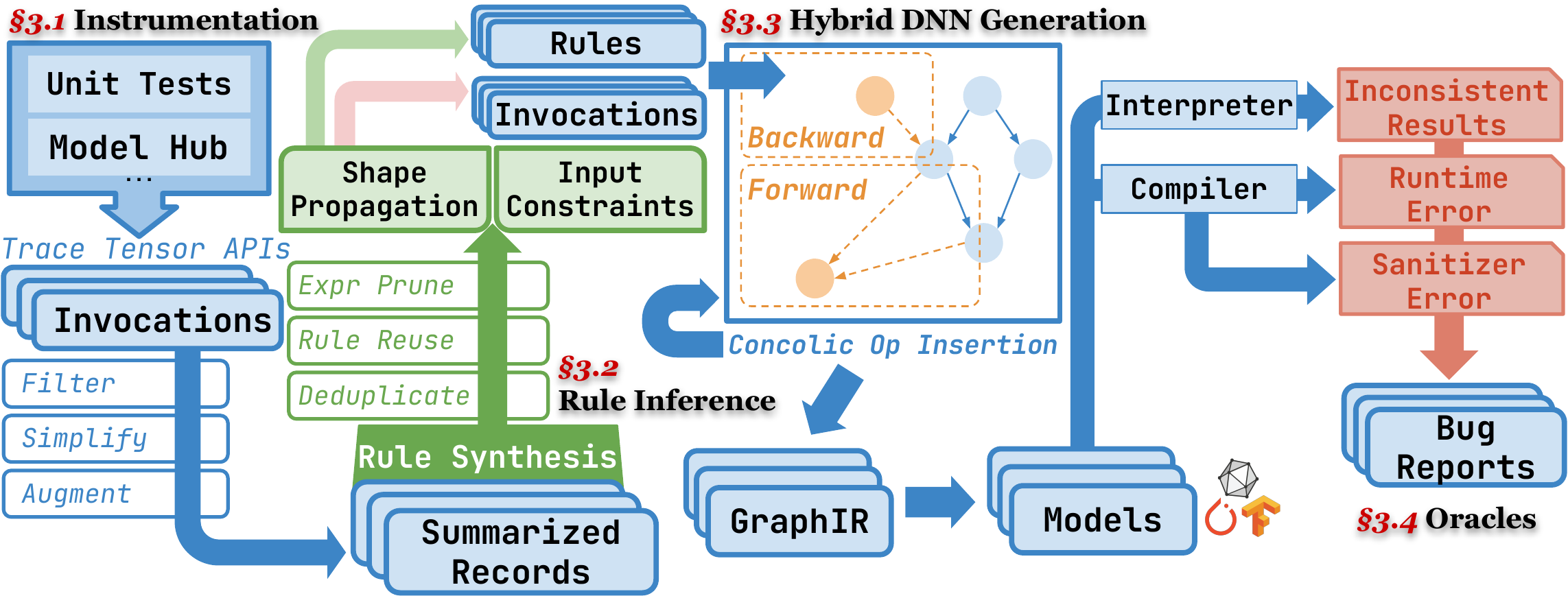}
    \caption{Overview of \sys}
    \label{fig:overview}
\end{figure*}

Figure~\ref{fig:overview} shows the overview of \sys's workflow.
\begin{itemize}
    \item 
    \sys improves the search space of model generation by making use of concrete invocation.
    To collect those \textit{desired} invocations of tensor APIs, we instrument various sources such as developer tests.
    Next, we filter out invocations that do not meet properties such as determinism, to facilitate rule inference and bug detection in later phases.
    For the convenience of rule inference, we summarize the invocation records to a simplified structure and further augment data diversity via mutation (\S\ref{sec:invoke}).
    \item 
    These records are discrete data points, 
    with which we can inductively synthesize arithmetic expressions in their corresponding operator rules.
    The inductive program synthesis problem is, however, an NP-hard~\cite{jha2017theory} problem, whose complexity rests with the grammar under enumeration.
    For affordability, we split a complete operator rule into multiple sub-rules, 
    in order to be describable by a simple arithmetic grammar.
    Furthermore, we prune the enumeration space over \emph{equivalence} and \emph{rarity}, and as a shortcut, reuse rules when possible.
    Additionally, redundant input constraints are removed for runtime efficiency.
    \item Next we apply \emph{Hybrid DNN Generation} which performs DNN generation over (i) symbolic operators, \ie those with operator rules; and (ii) concrete operators, \ie those whose rules are not inferred but with concrete invocation records.
    To achieve this, we perform \textit{concolic operator insertion} where symbolic operators are inserted with SMT solving while concrete ones are inserted by searching a compatible tensor type (\ie shape and data type).
    \item Lastly the generated model, after materialization, is cross-checked between the interpreter and compiler via oracles in \S\ref{sec:oracle}.
\end{itemize}

\subsection{Instrumentation}\label{sec:invoke}

\begin{figure}
    \centering
    \includegraphics[width=\linewidth]{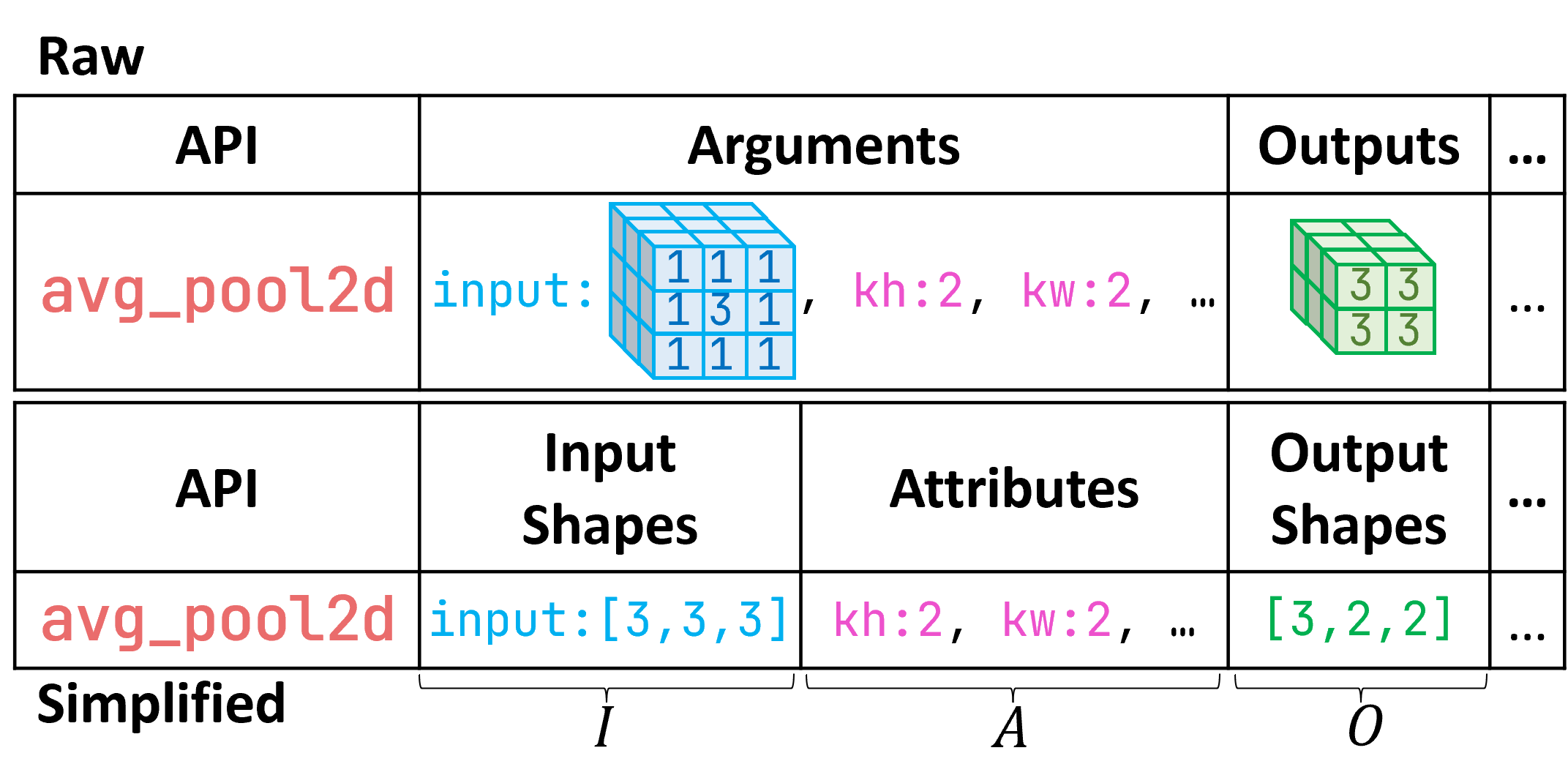}
    \caption{Layout of records before and after simplification.}
    \label{fig:record}
\end{figure}

\noindent\textbf{Invocation collection.}
Following prior work~\cite{wei2022free},
we instrument the desired APIs to store \textit{successful} invocations locally.
In contrast to \FreeFuzz{} which comprehensively instruments APIs from both tensor and high levels, we focus on \textit{tensor APIs} since high-level APIs can be decomposed to a series of tensor operations.
Additionally, these tensor APIs must be deterministic and value-independent (detailed in \S\ref{sec:impl}). 
We perform instrumentation at Python level (\eg over the Python test-suite) since Python is commonly used as the front-end of DL frameworks.
Meanwhile, we simplify the disorganized and superfluous raw invocation (Figure~\ref{fig:record}) by dropping concrete tensor values and thereby only preserve the tensor types, functor and other arguments.
For instance of \texttt{avg\_pool2d} (Figure~\ref{fig:rexamp}),
the layout of its simplified record\footnote{For clarity, we use ``record'' to represent ``simplified invocation records'' from now on.} is illustrated in Figure~\ref{fig:record}.

\noindent\textbf{Data augmentation via mutation.}
The robustness of inferred rules depends on the quantity and quality of records.
Unfortunately, by only using instrumented records, 
each rule on average only shares 5-7 records, which are insufficient.
More importantly, all collected records are \emph{passing examples}; however, \emph{counter examples} are also required for inferring input constraints (\S\ref{sec:inf}).
Consequently, we further diversify the records by mutating existing records,
where valid mutants are used as \emph{passing examples} (denoted by $\Records^\text{\cmark}$) and invalid ones are used as \emph{counter examples} (denoted by $\Records^\text{\xmark}$).
Specifically, we perform three phases of mutations over input shape dimensions and attributes of records (\ie $\attr \cup \dinp$):

\begin{enumerate}
    \item \emph{Offset-based}:
    The goal of offset-based mutation is to quickly build a large set of (preferably) passing examples.
    To achieve this,
    we enumerate subsets over $\attr \cup \dinp$,
    for each of which we increment the elements by 1
    until a desired number of records (\eg 100 in our experiments) or time budget runs out.
    The hypothesis behind is \emph{validity locality}: 
    oftentimes validity is preserved after a light-weight mutation (\ie increment by 1). 
    \item \emph{Swapping}:
    Exchanging two values from $\attr \cup \dinp$ can quickly verify simple inequalities.
    For example, assuming $a > b$ holds in all collected passing examples, we invalidate the inequality if the record is still valid after exchanging the values in $a$ and $b$.
    \item \emph{Special values}:
    Lastly, we randomly assign attributes with special values (\eg 0/-1) in order to test attributes' negativity.
\end{enumerate}

Meanwhile if no counter examples are produced after sufficient mutation,
it means probabilistically it has no (or extremely weak) input constraints.
Consequently, for counter-example-free operators, we directly assign an empty set as its input constraints (\ie no need to infer input constraints with inductive synthesis in \S\ref{sec:inf}).

\subsection{Rule Inference}\label{sec:inf}

We now explain how to perform rule inference via inductive program synthesis and make it affordable with a line of optimizations.

\noindent\textbf{Inductive synthesis of operator rules.}
Input constraints and shape propagation rules can be described by functions (over $\attr\cup\dinp$) whose bodies are oftentimes arithmetic expressions (\S\ref{sec:rule}).
Specifically, we can define such an arithmetic grammar $\gram$ as follows:

\newcommand{\decNT}[1]{{\sf #1}\xspace}

\vspace{-13pt}
\begin{bnf*}
  \bnfprod{\expr}
    {\bnfsp \bnfpn{\decNT{op}} \bnfsp \bnfpn{\expr} \bnfpn{\expr} \bnfor 
     \bnfpn{\decNT{item}}}\\
  \bnfprod{\decNT{op}}
    {\bnfts{+}\bnfor
     \bnfts{-}\bnfor
     \bnfts{$\times$}\bnfor
     \bnfts{$\div$}\bnfor
     \bnfts{min}\bnfor
     \bnfts{max}\bnfor
     \bnfts{mod}
     }\\
  \bnfprod{\decNT{item}}
    {\bnfpn{\decNT{symbol}}\bnfor
     \bnfpn{\decNT{constant}}
     }\\
  \bnfprod{\decNT{symbol}}
    {\bnftd{Symbols from $\dinp$ and $\attr$}}\\
  \bnfprod{\decNT{constant}}
    {\bnftd{Constant integers}}
\end{bnf*}

With the grammar,
we can infer an operator rule via inductive program synthesis, \ie by enumerating $\gram$ to find an expression that matches inputs/outputs of the records, in certain time budget and program size.
Because expressions in operator rules tend to be short,
we perform \emph{bottom-up} enumerative search~\cite{alur2017scaling,udupa2013transit} which first constructs small terms (\eg $\langle\decNT{item}\rangle$) and compose them gradually for generating larger ones.
For clarity we denote the set of enumerated expressions to be $\eset$.
Meanwhile, there are a few hypotheses to consider, \eg $|\attr|$, $|\dinp|$ and $|\dout|$ are assumed be fixed.
We will detail them later in the ``\ruleInst{}'' paragraph.

Formula~\ref{align:opi:sprop} describes a shape propagation rule with a set of expressions for computing corresponding output dimensions.
To infer the propagation expression for the i-th output dimension (\ie $o_i$), 
we enumerate $\eset$ until $\expr_k\in\eset$ is found to match all records (Formula~\ref{align:opi:spropmatch}).
Otherwise, we say the rules are not inferred and will not be used for inserting \emph{symbolic operators} during model generation.

\vspace{-5mm}
\begin{align}
\sprop \cong \{ o_1 = \expr{}_1(\dinp, \attr), \cdots, o_n = \expr{}_n(\dinp, \attr) ~|~ \expr{}_i \in \eset \}
\label{align:opi:sprop}
\end{align}
\vspace{-5mm}
\begin{align}
\exists\expr{}
\!
\in
\!
\eset, 
\forall \langle\insta{\attr},\insta{\dinp},\insta{\dout}\rangle\!\in\!\Records^{\text{\cmark}},
\expr{}\subst{\insta{\attr}\cup\insta{\dinp}}{\attr\cup\dinp}
\!
=
\!
o_i\subst{\insta{\dout}}{\dout}
\label{align:opi:spropmatch}
\end{align}

Similarly, input constraints $\icons$ are predicates of equalities and inequalities, which can be normalized to $0 = \expr$ and $0<\expr$ respectively.
Algorithm~\ref{algo:icons} illustrates how $\icons$, starting with an empty set (Line~\ref{algo:icons:emptyset}), is inferred.
By enumerating predicates oriented from $\eset$ within the time limit (Line~\ref{algo:icons:outer}-\ref{algo:icons:timeout}),
we find predicates that are satisfied by all passing examples (Line~\ref{algo:icons:precord}).
Specifically, if any passing example does not match the predicate under enumeration (\ie $c$) (Line~\ref{algo:icons:failpass}),
$c$ is then undesired, and consequently we restart the loop for the next predicate (Line~\ref{algo:icons:goto}).
Otherwise, we include $c$ in $\icons$ (Line~\ref{algo:icons:add}).
Meanwhile, for soundness input constraints should reject invalid inputs (if any).
Consequently,
$\icons$ should reject all (if any) counter examples (Line~\ref{algo:icons:counter}-\ref{algo:icons:passcounter}).
Otherwise, the rule is not inferred (Line~\ref{algo:icons:fail}).

\begin{algorithm}
\caption{Inference of input constraints $\icons$}\label{algo:icons}
\DontPrintSemicolon
\small
\SetKwProg{Fn}{Function}{:}{}
\SetKw{Continue}{continue}
\SetKw{Break}{break}
\SetKw{Raise}{raise}
\SetKw{Check}{check}
\SetKw{Foreach}{foreach}

\SetKwFunction{InferInputConstraints}{\textsc{InferInputConstraints}}
\SetKwFunction{Deduplicate}{\textsc{Deduplicate}}

\Fn{\InferInputConstraints{$\eset$, $\Records^\text{\cmark}$, $\Records^\text{\xmark}$}}{
    $\icons \gets\emptyset$ \label{algo:icons:emptyset}\;
    \OUTER:\ \For{$c \in \{0 = \expr, 0< \expr; \forall \expr\in \eset\}$}{\label{algo:icons:outer}
        \lIf{timeout}{\label{algo:icons:timeout} \Break }

        \For{$\langle\insta{\attr},  \insta{\dinp} \rangle\gets \Records^\text{\cmark}$}{\label{algo:icons:precord}
            \If{ $\texttt{evaluate}(c\subst{ \insta{\attr}\cup \insta{\dinp}}{A\cup I})$ is false }{\label{algo:icons:failpass}
                \Continue \OUTER\tcp*{Go to next $c$ at Line~\ref{algo:icons:outer}}\label{algo:icons:goto}
            }
        }

        $\icons \gets \icons \cup \{c\}$\label{algo:icons:add}\;
    }

    \For{$\langle \insta{\attr},  \insta{\dinp} \rangle\gets \Records^\text{\xmark}$}{\label{algo:icons:counter}
        \If{ $\texttt{evaluate}(\bigwedge \icons\subst{\insta{\attr}\cup \insta{\dinp}}{A\cup I})$ is true }{\label{algo:icons:passcounter}
            \Raise Inference failure\label{algo:icons:fail}
        }
    }

    \Return $\icons$\;
}
\end{algorithm}

\begin{figure}
    \centering
    \includegraphics[width=\linewidth]{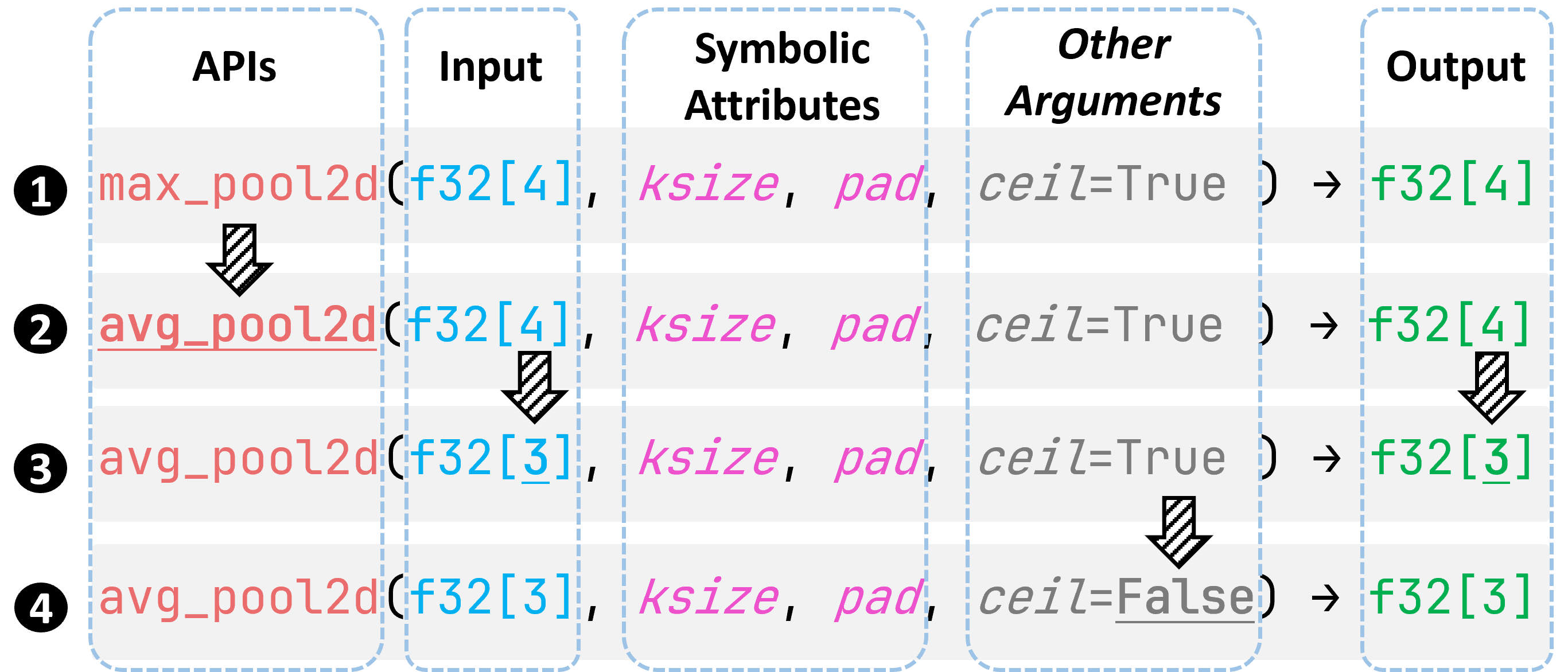}
    \caption{Examples of similar but distinct \ruleInsts{}. ``\texttt{f32[3]}'' stands for a float32 tensor variable whose rank is 3.}
    \label{fig:opins}
\end{figure}

\noindent\textbf{\RuleInst{}.}
In \nnsmith, operator rules are directly written in Python, whose grammar is much more complicated than $\gram$.
Running inductive program synthesis over such a complex grammar, though being more capable, is impractical.
To preserve a grammar as simple as $\gram$,
we split an operator into multiple \emph{\ruleInsts{}},
by ``fixing'' components whose variation incurs a more complicated grammar.
We can empirically summarize such components for defining \ruleInsts{}.
For example, branches in operator rules are used for handling variable lengths of inputs or input ranks.
Additionally, rules of operators with dimension-sensitive attributes, \eg \texttt{dim} in \texttt{max(x, dim)}, often requires array operations.
With Figure~\ref{fig:opins}, in addition to API names (\eg \myc{1} and \myc{2}), we identify a \ruleInst{} with the following properties:

\begin{enumerate}
    \item \emph{Tensor signature}:
    Recall \S\ref{sec:rule} that an operator could take and return a variable length of tensors in various ranks.
    Because incorporating such variability is costly, we let each \ruleInst{} have a fixed tensor signature (and thus a fixed form of $\dinp$ and $\dout$).
    For example, \myc{2} and \myc{3} have the same API name but are different \ruleInsts{} for having different input/output ranks (\ie 3 versus 4).
    It is also worth noting that we do not distinguish \ruleInst{} over the data types of input/output tensors which are often orthogonal to the operator rules. 
    \item \emph{Symbolic attributes}:
    Besides input tensors, we regard other arguments that can be symbolized to \textit{symbolic integers} as \emph{symbolic attributes}, which are the free variables in operator rules (\ie $\attr$ in \S\ref{sec:rule}).
    Therefore, invocations with different sets of symbolic attributes are associated with different \ruleInsts{}. 
    \item \emph{Other arguments}: 
    We further classify the rest of arguments (\ie non-tensor and non-symbolic-integer) into two categories: \textit{(i)} rule-orthogonal arguments (\eg float-point scalars such as ``bias'') and \textit{(ii)} (likely-)rule-dependent arguments (\eg image layout in ``NCHW'' or ``NHWC'').
    Only \textit{(ii)} will be used for identifying \ruleInsts{} for its potential impact on operator rules.
    In general, the sub-category of an \emph{other argument} is determined by its type and value.
    For instance, in Figure~\ref{fig:opins}, the \texttt{ceil} argument, as a boolean, falls into the \textit{(ii)} category, which makes \myc{3} and \myc{4} different \ruleInsts{}.
    In fact, \texttt{ceil} impacts the shape propagation rule of \texttt{avg\_pool2d}, where being true makes the output height and width rounded by \emph{ceil} instead of \emph{floor} (see Formula~\ref{eq:poolshape}).
    Further details will be elaborated in \S\ref{sec:impl}.
\end{enumerate}

\noindent\textbf{Pruning.}
$\eset$ can be too large to enumerate. %
For efficiency, we prune semantic-equivalent duplicates (\ie \emph{equivalence}), as well as those that are uncommon in operator rules (\ie \emph{rarity}).
Specifically, we list the pruning methods in their order of being applied:

\newcommand{\constexpr}{$\langle op\rangle\  \langle constant\rangle\ \langle constant\rangle$\xspace}
\begin{enumerate}
\item \emph{Bound}:
    Without constraints, $\eset$ is infinite. 
    Therefore, we bound $\eset$ by limiting the number of $\langle op\rangle$ and the set of constant literals.
    For example, in our default experimental setting the maximum number of $\langle op\rangle$ being used is 5 and we use $\langle constant\rangle\gets \{1, 2\}$.
    Additionally, expressions describing inequality are further limited to have at most one $\langle op\rangle$ because \emph{(i)} inequalities are oftentimes simple;
    and \emph{(ii)} a larger upper limit will lead to many false-positives as inequalities are more flexible than equalities.
\item \emph{Rarity}:
    We empirically prune expressions with the same symbols occurring more than once, which are uncommon.
    Those with constant sub-expression (\ie \constexpr{}) are also pruned for being constant foldable.
\item \emph{Equivalence}:
    We find semantically equivalent expressions in $\eset$ and only keep the simplest one.
    Specifically, we leverage a two-pass approach inspired by Ruler~\cite{ruler}:
    First, for the expressions with the same free variables, we \textit{quickly} evaluate them over a number of randomly generated assignments and group them according to the outputs (\ie often known as \textit{characteristic vectors} or \textit{finger-prints}).
    For each group, we then rigorously find equivalents by applying an SMT prover.
\end{enumerate}

$\eset$ may vary for different $\attr\cup\dinp$.
While pruning $\eset$ for each \ruleInst{} is costly,
we make it one-time effort by:
(i) pruning $\eset_\square$ with ``holes'' (\ie symbol placeholders);
and 
(ii) extending $\eset_\square$ to $\eset$ by replacing the holes with actual symbols of each \ruleInsts{}.
For example,
assume that $\eset_\square$ is the pruned set of expressions with holes of $\{\square_1,\square_2,\square_3\}$, \ie $\langle\decNT{symbol}\rangle\ ::=\ \square_1\ |\ \square_2\ |\ \square_3$.
To infer an operator rule with $\attr\cup\dinp\gets\{s_1,\cdots,s_4\}$, we get the actual $\eset$ by extending $\eset_\square$, 
by mappings $\{s_1,\cdots,s_4\}$ to $\{\square_1,\square_2,\square_3\}$ in various ways.
More specifically,
for each $\expr$ with $h$ holes (symbols),
we select $h$ symbols from $\attr\cup\dinp$ to ``fill'' the holes (\ie substitution).
Because of the \emph{one-time-occurrence} hypothesis, each symbol in $\attr\cup\dinp$ will not be selected to fill multiple holes (\ie injective).
In addition, the mapping from the selected symbols (in $\attr\cup\dinp$) to holes is determined according to the relative order of indices.
For example, for $\square_1 \div \square_2$, by selecting $\{s_1, s_2\}$ we only get $s_1 \div s_2$.
Why not consider permutation over the mappings?
Consider if we allow $s_2 \div s_1$ as an extension, 
when extending $\square_2 \div \square_1$ (indices swapped) we get the same duplicated expression.
As a result, for each $\expr$ with $h$ holes, we can extend ${|\attr\cup\dinp| \choose h}$ expressions.
Meanwhile, when $|\attr\cup\dinp|$ is smaller than the maximum number of holes, we only consider extending $\expr$ where $h \le |\attr\cup\dinp|$.

\noindent\textbf{Rule reusing.}
\RuleInsts{} can share equivalent rules.
Before running inference from scratch, we can first test if the records can be matched by already inferred rules (if they share the same form of $\attr$, $\dinp$ and $\dout$). %
If an existing rule can be matched, we can simply ``copy and paste'' it for the new \RuleInst{} as a short-cut; otherwise, we can still run inference from scratch.
Furthermore, since this optimization is orthogonal to the grammar, we can also reuse those expert-crafted rules from \nnsmith (Python grammar).

\begin{algorithm}
\caption{Predicate deduplication for $\icons$}\label{algo:dedup}
\DontPrintSemicolon
\small
\SetKwProg{Fn}{Function}{:}{}
\SetKw{Continue}{continue}
\SetKw{Break}{break}
\SetKw{Raise}{raise}
\SetKw{Check}{check}
\SetKw{Foreach}{foreach}

\SetKwFunction{Deduplicate}{\textsc{Deduplicate}}

\Fn{\Deduplicate{$\icons$}}{
    \Repeat{$\icons$ unchanged}{
      \For{$c \in \icons$}{
            \If{\texttt{Prove}$(\bigwedge \left[\icons\right] \Leftrightarrow \bigwedge \left[\icons  - \{c\}\right])$}{\label{algo:dedup:prove}
                $\icons \gets \icons  - \{c\}$\;\label{algo:dedup:rm}
            }
        }
    }
}
\end{algorithm}

\noindent\textbf{Deduplication.}
The inferred input constraints could have many redundant predicates, which slows down SMT solving when fuzzing online and makes it less readable.
Therefore, Algorithm~\ref{algo:dedup} deduplicates the predicates obtained from Algorithm~\ref{algo:icons}.
We each time remove a predicate $c$ if $\icons$ is equivalent to that without $c$ (Line~\ref{algo:dedup:prove}-\ref{algo:dedup:rm}).
We run the algorithm until a fixed point when no predicates from $\icons$ are removed after an iteration.

\subsection{Hybrid DNN Generation}\label{sec:gen}

Following the Algorithm 1 in \nnsmith~\cite{nnsmith}, the generation of a DNN can be regarded as a problem of how to insert a \emph{valid} operator \textit{correctly} to an already \textit{valid} DNN model.
In \nnsmith, since all operators has their corresponding rules (implemented by domain experts), a DNN model is synthesized \emph{symbolically}, \ie all shape dimensions and attributes are viewed as symbols at construction time and later materialized with a set of assignments offered by the SMT solver.
For \sys, we adopt a \emph{concolic}~\cite{sen2005cute} style of DNN generation in order incorporate both symbolic and concrete operators\footnote{A \emph{symbolic operator} is an operator with inferred rules; while a \emph{concrete operator} does not have rules successfully inferred but still has its corresponding validated records.}.

Specifically, the hybrid DNN generator inserts operators concolicly such that each operator, after insertion, is \textit{immediately} concretized by the model from the solver, instead of deferring it when a full DNN is built.
Therefore, the DNN under construction is always concrete, \ie all shape dimensions, data types and attributes are concrete at construction time, which makes it applicable to insert a concrete operator.
There are a few benefits with a concrete DNN:
\emph{(i)} the insertion of concrete operators can be efficiently implemented by looking up compatible tensor types between traced records and the model under construction;
and
\emph{(ii)} theoretically SMT solving is incurred less intensively (thus faster) with less symbols.
In \nnsmith, an operator can be inserted in two directions: 1) \textit{forward insertion} that inserts an operator which consumes existing values; and 2) \textit{backward insertion} that lets an operator be a producer by occupying existing placeholders.
Next, we elaborate how symbolic and concrete operators are \textit{forward} inserted and for clarity omit the details for backward insertion, which can be regarded as a reversed version over the placeholders (instead of arbitrary tensors).

\noindent\textbf{Inserting a symbolic operator.}
Both the manually written operator rules (\ie \nnsmith) and inferred ones (\ie \sys) can be used to insert symbolic operators.
Operators in both groups are selected separately in equal amount of probability.
Each time, to insert a selected operator $\op$, we first enumerate arity-sized combinations of tensor variables as the input candidates to $\op$, 
where each of them must respect the data type and rank requirements of $\op$.
Taking the batched 2D-convolution as an example, whose input tensor must have four dimensions, any other tensors whose rank is not four will not be taken into the enumeration.
Next, each of the input candidate tuples is checked by the input constraints $\icons_\op$ until one satisfies $\icons_\op$ and thus becomes the tensor inputs $\rinp$ to $\op$.
Furthermore, we ask the SMT solver to provide a model from the input constraints and use the assignments of operator attributes $\insta{\attr}$ to initialize $\insta{\op} \gets \op\subst{\insta{\attr}}{\attr}$.
We then insert $\insta{\op}$, taking $\rinp$ as inputs, to the DNN under construction.
We also propagate its output tensor types with the shape propagation rule (Formula~\ref{align:opi:sprop}) for making future insertion feasible (\S\ref{sec:rule}).
Of course, if none of the candidates can make the rule satisfiable, the insertion of $\op$ will be discarded and the algorithm will re-try another operator.

\noindent\textbf{Inserting a concrete operator.}
The feasibility of inserting a concrete operator is as simple as finding an intersection of tensor types between inputs (\ie $\dinp$) in records and visible variables in the working DNN.
For example, 
given a \texttt{avg\_pool} which has input tensor type of \texttt{float32[1,3,224,224]} in the records, 
if there is a tensor variable with a shape of \texttt{[1,3,224,244]} and a data type of \texttt{float32} in the DNN under construction,
we can safely insert it to the target place.
However, because of the large volume of records, checking the satisfiability operator by operator and record by record is inefficient.
Instead, we can build a mapping from tensor type (\ie shape plus data type)
to a set of \ruleInsts{}, any of which has an input tensor of such a type.
Then, by accumulating the set of \ruleInsts{} mapped from tensor types available in the working DNN, we get a reduced set of operator candidates which exclude those with unsatisfiable input types. %
Consequently, we only need to enumerate records of reduced sets of \ruleInsts{}.
Once a record if found to be matchable, we initialize the \ruleInsts{} with the record and insert it to the DNN under construction.

\subsection{Test Oracle}\label{sec:oracle}

In this section, we list three test oracles for manifesting bugs.

\noindent\textbf{Result inconsistency.}
In addition to running the DNNs \emph{eagerly} with the pre-compiled library functions (\ie \emph{interpreter}),
TensorFlow and PyTorch can further optimize the models via \emph{compilation} for better performance.
Hence, 
we cross-check the results obtained by running the same model and inputs from the \emph{interpreter} and \emph{compiler},
where a subtle floating-point error is allowed.

\noindent\textbf{Runtime error.}
We identify a runtime-error bug if the compilation or execution of a model aborts unexpectedly.
The corresponding symptoms include a crash or an unexpected Python exceptions not incurred by incompatibility (\eg  ``Not-implemented'' error).
Furthermore, interpreter exceptions are not considered as bugs as it could be caused the use of an incorrect operator rule.

\noindent\textbf{Sanitizer error.}
We also enrich the test oracles
with sanitizers, such as ASan~\cite{asan} (memory error), UBSan~\cite{ubsan} (use of compiler's undefined behavior), and CSan~\cite{csan} (CUDA error).
Sanitizer errors are reported by sanitizer-injected checkers at runtime.
Without sanitizers, bugs may not manifest themselves via a crash (\eg buffer overflow) or occur at a late stage, making debugging challenging. 

\section{Implementation}\label{sec:impl}

\sys implements three components including an instrumentation tool and rule synthesizer (\ie offline), as well as a fuzzer (\ie online), through an effort of \locTotal LoC in total.

\noindent\textbf{API instrumentation.}
The instrumentation tool is implemented with \locInstrumenter LoC in Python.
The instrumentation is performed by inserting an API-hijacking code snippet in the \texttt{\_\_init\_\_} files of DL framework packages.
Specifically, we run the instrumentation over the developer tests from the open-source repositories of PyTorch and TensorFlow.
As soon as the DL packages are imported, 
the API-hijacking code adds a function wrapper to all functions within the package.
During execution (\ie running the regression tests of DL frameworks), the invocation snapshots to desired APIs are serialized for reproduction.
In post-processing, a filter is applied to remove invocations that do not comply with \textit{determinism} and \textit{value independence}.
To detect \textit{determinism}, each invocation is replayed for three times for checking output consistency.
For testing \textit{value independence}, \ie the operator rules are independent to the values/elements in the input tensor,
each API is tested by three groups of random inputs (initialized from $-10^6$ to $10^6$) and is expected to output the same tensor types without runtime failure. 
This component also includes utilities for parsing and composing/replaying a DL API, in order to construct new invocation.

\noindent\textbf{Rule synthesizer.}
We implemented the synthesizer in \locSynthesizer{} LoC in Python.
Before the actual rule synthesis, we first apply data augmentation for enriching the records on demand (\ie until 100 records for each \ruleInst{}).
In \S\ref{sec:inf} it is found that not all arguments impact the rules,
consequently we identify \emph{rule-orthogonal} arguments in a \ruleInst{} through the argument type: for a floating-point argument (\eg bias) we assume it does not impact operator rules.
Furthermore, the arithmetic expressions are represented as binary trees and for memory efficiency smaller trees are re-used to compose larger trees via pointers in the bottom-up enumerative search.
We constrain the maximum number of $\langle\decNT{op}\rangle$ to 5 (\ie 6 symbols at most).
As a result, as a one-time effort we first enumerate $\gram$ by regarding the symbols as 6 ``holes'' and prune it on the fly to get $\eset_\square$.
Meanwhile, we also leverage commutativity of $\{+, \times,\min,\max\}$ to skip the enumeration of operand swapping and associativity over $\{+,-, \times, \div, \min, \max\}$, in order to accelerate the equivalence-based pruning.
With $\eset_\square$ obtained as a one-time effort,
for any new operator rule under inference,
we can quickly extend $\eset_\square$ to $\eset$ by filling its actual symbols into the ``holes'' as discussed in \S\ref{sec:inf}.
Specifically, for each \ruleInst{} we set a timeout of 1000 seconds to infer the shape propagation or input constraints. 

\noindent\textbf{Fuzzer.}
The fuzzing engine of \sys is built by extending the \nnsmith{} prototype~\cite{nnsmithartifact} with \locFuzzer new LoC (and removing \locRmFuzzer{} old LoC).
Major efforts are spent to improve the extensibility and debuggibility of the original \nnsmith{} for benefiting algorithm prototyping and bug finding.
Previously \nnsmith{} uses directed multi-graphs in \texttt{networkx}~\cite{networkx} for describing DL models internally.
However, the graph data structure is not suitable for manipulating model structures and being translated to real-world model formats.
Additionally, DL models are fundamentally programs~\cite{tf-security-program} which is not necessarily always pure data-flow graphs.
For example, in-place operators for reproducibility require a total order during execution whereas traversing a graph cannot guarantee.
As a result, we build an SSA-based intermediate representation, namely \gir, to describe DNN structures.
Following the LLVM~\cite{lattner2004llvm} interface style, DNN manipulation is made safe and convenient via three fundamental APIs of \texttt{insert}, \texttt{remove\_unused}, and \texttt{replace\_alluse}.
Thanks to the extensibility, 
five graph generation strategies used in this paper,
including three \sys variants and two \nnsmith{} variants,
are implemented in merely \locFuzzerCore{} LoC.

\section{Evaluation}

We evaluate \sys by asking following research questions:

\begin{itemize}
    \item \textbf{RQ1 (\S~\ref{sec:rq1})}: How does \sys compare against state of the art in DL compiler fuzzing in terms of code coverage?
    \item \textbf{RQ2 (\S~\ref{sec:rq2})}:
    How many APIs, \ruleInsts{} and records are collected and eventually inferred with operator rules?
    How efficient and effective is our rule synthesizer compared with general-purpose program synthesis tools such as \Rosette~\cite{rosette}?
    \item \textbf{RQ3 (\S~\ref{sec:rq3})}: How effective is \sys when detecting previously unknown bugs for real-world DL compilers?
\end{itemize}

\subsection{Experimental Setup}

\newcommand{\tba}{\textsf{T1}\xspace}
\newcommand{\tbb}{\textsf{T2}\xspace}
\newcommand{\torchjit}{PyTorch JIT\xspace}

\noindent\textbf{Systems under test.}
We test the emerging \textit{compilers} of the most popular DL frameworks, \ie TensorFlow~\cite{tensorflow} and PyTorch~\cite{pytorch}, which for clarity are denoted by ``TF'' and ``PT'' respectively.

\begin{enumerate}
    \item \textit{TensorFlow XLA} compiler converts a TensorFlow model (\eg SavedModel) to its graph-level IR (\ie HLO) for running various optimization passes. 
    TensorFlow defines over 1500 operators.
    Of these, around 450 are supported by XLA.
    This is because DL compilers often focus on a small set of primitive operators, from which other high-level operators can be composed.
    \item \textit{\torchjit{}}, the PyTorch's equivalent of XLA, supports around \jitops{} APIs (including alias, \eg \texttt{torch.max(a)} and \texttt{a.max()}) out of a total of over 2000 PyTorch operators.
\end{enumerate}

\noindent\textbf{Metrics}.
We evaluate \sys over various metrics.
Specifically, we explain the most important two here and defer the others.

\begin{itemize}
\item \emph{\# Found bugs}:
    We count the bugs at the basis of bug reports, which are classified to four statuses:
    1) \emph{fixed}: A patch has been effectively applied to fix the bug;
    2) \emph{confirmed}: 
    In addition to fixed bugs,
    we \emph{conservatively} (\ie lower bound) identify a confirmed bug \emph{iff} it has been reproduced/diagnosed as a fault or directly assigned to developers for fixing it;
    3) \emph{won't fix}: Developers claim the potential of not fixing it (\ie upper bound); 
    and 4) the rest of bugs are all triaged but require further investigation.
\item \emph{Branch coverage}:
    Following~\cite{tzer,nnsmith}, we evaluate fuzzers with \textit{branch} coverage,
    a stronger criterion (than line coverage) for test adequacy~\cite{testadequacy},
    over DL frameworks' C++ source code.
\end{itemize}

\noindent\textbf{Baselines.}
In end-to-end benchmarks, we compare \sys with the state-of-the-art model-level fuzzers (namely \nnsmith{}~\cite{nnsmith} and \muffin{}~\cite{muffin}) and the state-of-the-art operator-level fuzzer (\ie \deeprel{}~\cite{deeprel}).
For ablation study we also evaluate \sys's variants.

\begin{itemize}
    \item \nnsmith{} performs model generation with over 60 operators whose rules are \emph{manually} crafted by domain experts.
    Specifically, the official \nnsmith{} performs \emph{pure symbolic} generation where the all symbols are materialized together when all operators in the graph are symbolically inserted.
    In this paper we also propose concolic generation (\S\ref{sec:gen}), consequently we also implemented a concolic version of \nnsmith{} which immediately materializes the symbols per insertion. 
    Because the concolic variant performs similarly as the  pure-symbolic version, for clarity we omitted the results in evaluation.
    One reason can be that concolic insertion does not bring more operator supports in \nnsmith{}  as \sys{}.
    \item \muffin{} based on 11 seed models including DenseNet~\cite{densenet} and LSTM~\cite{lstm}, performs mutation-based model generation.
    Similar to \nnsmith{}, it supports over 60 operators by hand-crafting the shape inference rules.
    However, \muffin{-created} models are not guaranteed to be valid for the lack of input constraints.
    Notably, \muffin{} is only implemented on TensorFlow. As a result, we only compare \muffin{} against others on TensorFlow.
    \item \deeprel{} is an operator-level mutation-based fuzzer similar to \freefuzz{}.
    As an improvement to \freefuzz{} whose seeds purely come from instrumentation, \deeprel{} extends the seed invocations by matching similar APIs and exchanging their arguments.
    \item \sysr{} is a variant of \sys where the use of inferred operator rules is disabled.
    In other words, \sysr{} constructs DNNs by either using concrete operators determined by collected records or symbolic operators from the original \nnsmith{} and both methods share equal probability for being selected.
    \item The \sysi{} variant disables concrete operators and only uses symbolic operators from automated inference or \nnsmith{}.
\end{itemize}

In addition, in RQ3 we also compare our rule synthesizer with \Rosette, a solver-aided programming system, which supports inductive program synthesis.
Specifically, we give \Rosette $\gram$ as the grammar under a bit vector theory.
The number of bits for the data and operations is 32 given that the maximum number in records is $2^{32}-1$ (\ie \texttt{INT\_MAX}).
Next, for each \ruleInst{} we let its records be the constraints and run \Rosette with a 1000-second time budget.
More specifically, we only compare with \Rosette over the inference of shape propagation, \eg given an \ruleInst{} with k output dimensions, both \Rosette and \sys will run k times each of which trying to search $\expr_k$ that matches records of $o_k$.
We did not infer input constraints for \Rosette since multiple matched predicates can be returned in one pass while \Rosette directly terminates when the first matched predicate is found.

\begin{figure*}[h]
    \begin{subfigure}{0.33\textwidth}
        \includegraphics[width=\linewidth]{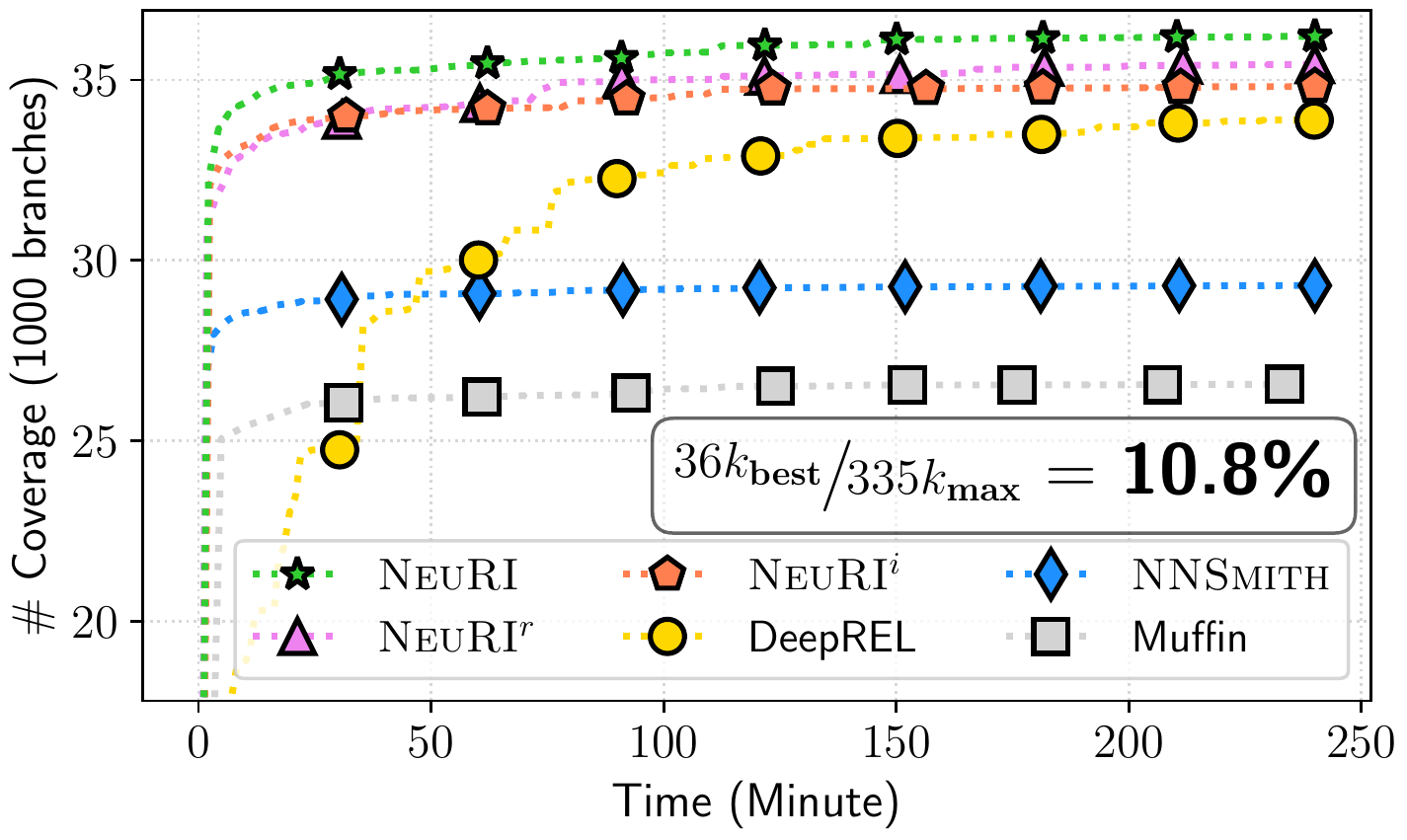}
        \caption{\eTF{}}
    \label{fig:covtf}
    \end{subfigure}
    \begin{subfigure}{0.33\textwidth}
        \includegraphics[width=\linewidth]{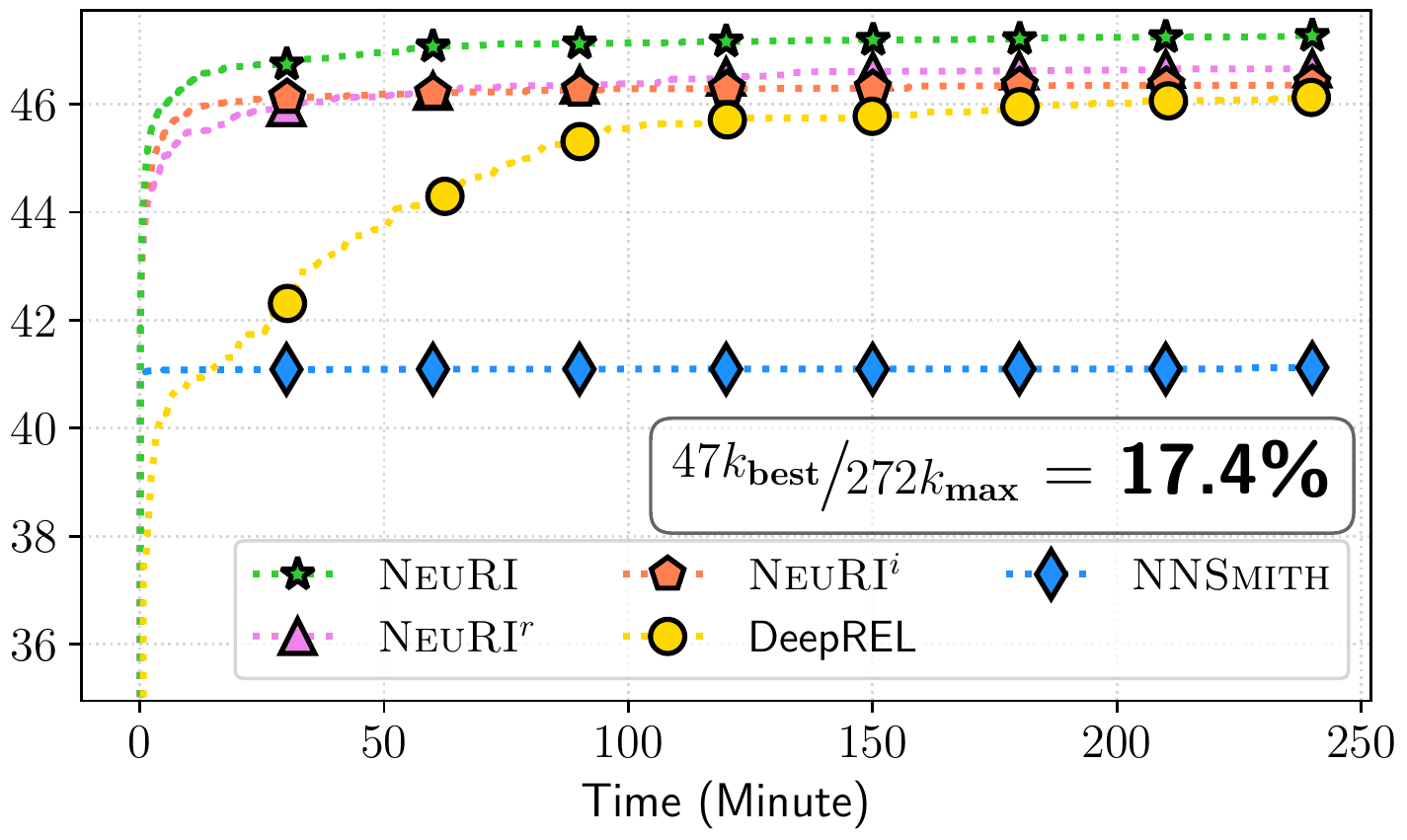}
        \caption{\ePT{}}
    \label{fig:covpt}
    \end{subfigure}
    \begin{subfigure}{0.33\textwidth}
        \includegraphics[width=\linewidth]{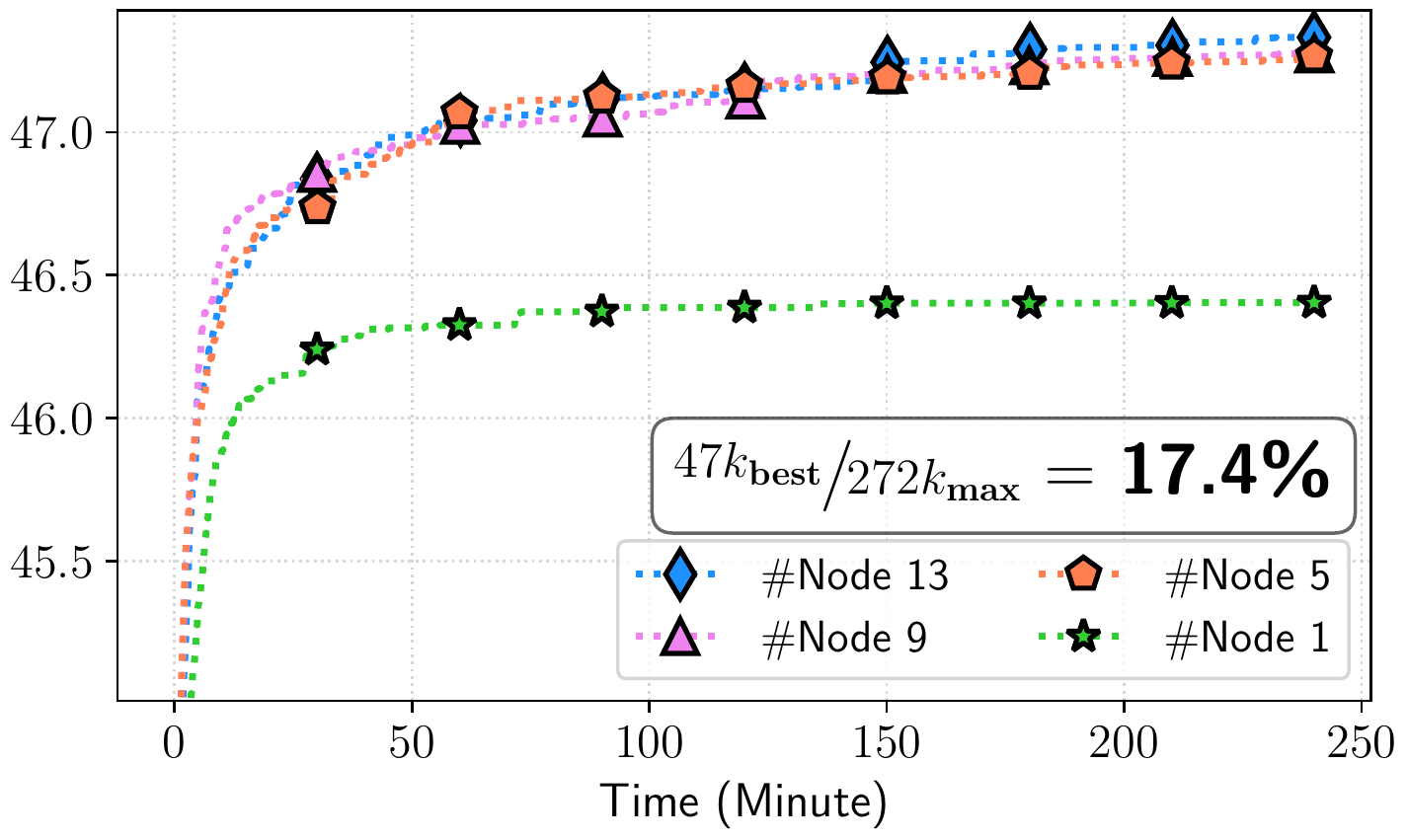}
        \caption{Impact of model size.}
    \label{fig:covnsize}
    \end{subfigure}
    \caption{4-hour coverage trend of fuzzing.}
\end{figure*}

\noindent\textbf{Configuration.}
We run all experiments on Ubuntu 22.04 powered by a 64-thread AMD Threadripper CPU, 256 GB of memory, and 4 TB of PCIe-4 SSD.
Our approach is evaluated over the up-to-date frameworks and versions:
TensorFlow v2.12-nightly (git: \texttt{5a6fc06bf8}) and PyTorch v2.1-nightly (git: \texttt{f7520cb51e}).
Due to the different tool-chain flavours, 
we compiled TensorFlow with GCC-12.2 and GCOV~\cite{gcov}, 
while PyTorch is compiled with Clang-14 and its source-code based coverage tool~\cite{llvmcov}.
To precisely measure the test adequacy of compilation,
for TensorFlow we instrument files in \texttt{tensorflow/compiler} (over 800k LoCs), and for PyTorch, non-kernel-function files under \texttt{pytorch/csrc} and \texttt{aten/} are conservatively instrumented since PyTorch's passes are ``everywhere''.
Following prior work~\cite{nnsmith,tzer} we by default run fuzzing for four hours and generate models with five operators to balance between efficacy and debuggability (Figure~\ref{fig:covnsize}).
For detecting result inconsistency, our oracle uses absolute error of $10^{-3}$ and relative error of 1\%.
With models generated as small as five nodes, we did not see false-positives brought by propagated floating-point errors.

\subsection{RQ1: Evaluating Coverage}\label{sec:rq1}

\noindent\textbf{Overall coverage.}
Figure~\ref{fig:covtf} and~\ref{fig:covpt} show the coverage growth (y-axis) in four hours (x-axis).
Among all \emph{model-level} fuzzers, \sys improves the prior SOTA (\ie \nnsmith{}) by \CovTFImproveModel{} on \eTF{} and \CovPTImproveModel{} on \ePT{}.
The results indicate that \sys can synthesize more diverse model structures to exercise various compiler passes.
In addition, \sys also outperforms the SOTA \emph{operator-level} fuzzer \deeprel{} by \CovTFImproveOp{}
on \eTF{} and \CovPTImproveOp{} on \ePT{}.
Though existing operator-level fuzzers~\cite{wei2022free,deeprel,xie2022docter} is known for constructing invocations for thousands of APIs, \sys shows that by only supporting a smaller but essential set of operators, similar and even better code coverage can be achieved through model-level generation. 
Despite the marginal coverage improvement over \deeprel{}, \sys, via model-level fuzzing, can explore various system behaviors that are not covered by \deeprel{}.
Specifically, \ppercent{39}{\PTBugFuzzing{}} of the \ePT{} bugs and \ppercent{5}{14} of \eTF{} bugs found by \sys are manifested by models of multiple operators which cannot be covered by single-operator fuzzers like \deeprel{}. 
Overall, 
in four hours \sys automatically covered \CovTFPerc{} and \CovPTPerc{} of the compiler code branches in \eTF{} and \ePT{} respectively.
It is worth noting that achieving 10-18\% of \emph{total branch} coverage is non-trivial, given the complexity of DL frameworks.
As a reference, fuzzers~\cite{kim2020hfl} for Linux (\ie also a million-LoC project) commonly achieve around 10\% of block coverage.
The missing branches can come from passes designed for other hardware targets (\eg GPU whereas CPU is used in our experiments) and unused experimental compilation pipelines (\eg MLIR).

\noindent\textbf{Ablation study.}
By comparing \nnsmith{} with \sysr{} and \sysi{},
we show that both symbolic and concrete operators are effective for improving coverage 
($18.8 \sim 20.9\%$ for \eTF{} and $12.8 \sim 13.5\%$ for \ePT{}).
By combining both of them together (\ie \sys), \sysr{} and \sysi{} can be further improved to cover $2.2\%$ / $1.3\%$ more branches for TF / PT.
Though this improvement might look marginal, we argue that the additionally covered branches are harder-to-reach ones, as we later show that a certain number of bugs are exclusively contributed by inferred rules.
Table~\ref{tab:valid} shows that \sys has a slightly lower validity rate ($<5.2\%$) and runs $17.6\%$ (TF) / $67.7\%$ (PT) slower than \nnsmith{}.
The validity rate is lower as some inferred rules are partially correct. 
The speed of \sys is of course slower as its model generation tackles over 75$\times$ more symbolic operators than \nnsmith{} and even much more concrete operators.
However, \sys still achieves better coverage as it generates test-case of higher quality over quantity.
Note that an non-prefect validity rate does not introduce false-positives, 
as we identify and immediately discard an invalid test-case if it raises exceptions in eager mode.

\noindent\textbf{Speed.}
The time cost for generating, evaluating, and storing a test-case is presented in Table~\ref{tab:tstat}.
It indicates that constrain-based model generation is efficient, 
 taking an average of 88ms (TF) / 69ms (PT) for creating and running a test-case, despite the single-thread nature of our model generator.
Specifically,
test-case generation on average takes 53\% (TF) / 86\% (PT) of the time,
with constraint solving accounting for about half of it.
Notably, the SMT solving time is long-tailed:
the P99 shows that in 1\% of the cases, the solver time deteriorates by over $12.5\times$ (TF) / $15.5\times$ (PT) compared to the average.

\noindent\textbf{Impact of model size.}
\sys and \nnsmith{} by default generate models with 5 nodes.
How about other sizes?
Figure~\ref{fig:covnsize} shows that the model size used in \sys impacts the coverage on \ePT{} (\eTF{} is omitted for clarity but shares similar trends).
For example, only generating one-operator models (\ie single-API) gets worse coverage, since compiler passes look for multiple-operator patterns (\eg operator fusion).
On the contrary, the benefits converges when model size gets larger and larger (\eg 0.1\% coverage difference in \# node 5-13).
Consequently, because smaller models are easier to diagnose, we by default use a model size of 5 in \sys.

\begin{table}
\caption{Number of valid tests generated in 4 hours.}
\label{tab:valid}
\centering
\begin{tabular}{c
c
>{\raggedleft\arraybackslash}p{1cm}
c
>{\raggedleft\arraybackslash}p{1cm}
}
\hline
     &  \multicolumn{2}{c}{TensorFlow}
     &  \multicolumn{2}{c}{PyTorch}    \\
     \cmidrule(lr){2-3} \cmidrule(lr){4-5}
     & \%Validity & \# Tests & \%Validity & \# Tests \\
\hline

\sys{}     &  94.8\% & 108,572  & 98.9\% & 206,486 \\
\sysi{}    &  94.2\% & 78,083   & 98.6\% & 134,912  \\
\sysr{}    &  98.1\% & 125,059  & 99.8\% & 409,872 \\
\hline
\nnsmith{} &  100\%  & 131,799  & 100\%  & 639,434 \\
\muffin{}  &  95.9\% &    302   &  ---   &    ---  \\

\hline
\end{tabular}
\end{table}

\begin{table}
\caption{Testing time breakdown (millisecond).}
\label{tab:tstat}
\centering
\begin{tabular}{l c cccc}
\hline
     &              & Gen. (SMT)     & Eval.      & Save        & Total  \\
\hline

\multirow{4}*{TF} 

& Avgerage   & 67 (33)   & 21 & 38 & 126 \\
& P90        & 74 (37)   & 29 & 42 & 145 \\
& P99        & 673 (411) & 58 & 59 & 733 \\
\cmidrule(lr){2-6}
& Percentage & 53\%   (26\%  ) & 17\%   & 30\% & 100\% \\
\hline

\multirow{4}*{PT} 
&Avgerage   & 60 (38) & 9 & 0 & 70 \\
&P90        & 45 (26) & 11 & 0 & 56 \\
& P99        & 930 (631) & 16 & 1 & 942 \\
\cmidrule(lr){2-6}
& Percentage & 87\%   (55\%  ) & 13\%   & 0\% & 100\% \\

\hline
\end{tabular}
\end{table}

\subsection{RQ2: Evaluating Rule Inference}\label{sec:rq2}

\noindent\textbf{Statistics.}
Table~\ref{tab:dist} displays the statistics of APIs, \ruleInsts{} and records at different phases.
The ``Collected'' row indicates that
the developer tests (\ie the instrumented code) incorporate 758 (out of \jitops{}) APIs supported by \torchjit{} and 248 (out of \xlaops{}) for XLA respectively.
42-45\% APIs are not collected due to the lack of tests (\eg untested aliased APIs) or being non-tensor APIs (\eg image encoder and decoder APIs).
By focusing on these, around 63k / 34k records can be collected for \ePT{} / \eTF{}.
After filtering out unwanted records (\S\ref{sec:impl}),
47\% (PT) / 38\% (TF) of the records remained for 90\% (PT) / 86\% (TF) of the APIs.
Furthermore, data augmentation improves the unique records by $15\times$ (PT) / $7.7\times$ (TF),
out of which 57\% (PT) / 67\% (TF) are counter examples.

Within the 1000-second budget per rule,
\sys can infer 76\% (PT) / 84\% (TF) of rules at the \ruleInst{} level and 91\% (PT) / 90\% (TF) of rules at the API level (Table~\ref{tab:dist}).
To estimate the usefulness of rules, we use ``fuzzing$^\top$'' to denote the number for APIs that are used during fuzzing (\ie \sysi in one node).
To indicate correctness, ``fuzzing$^\bot$'' denotes those in ``fuzzing$^\top$'' that always construct valid usages.
It turns out 97\% (PT) / 90\% (TF) APIs out of the inferred ones can be \textit{used} and 96\% (both) tend to be used \textit{validly}, showing the overall effectiveness of rule inference. 

\newcommand{\TableTorch}{\small PT\xspace}
\newcommand{\TableTF}{\small TF\xspace}
\newcommand{\TabMCLen}{1.7cm}
\newcommand{\TabMCSubLen}{0.65cm}

\begin{table}
\caption{\# API/\ruleInst{}/record at different stages.}\label{tab:dist}
    \centering
    \begin{tabular}{l rrrrrr}
    \hline
           & \multicolumn{2}{>{\centering\arraybackslash}p{\TabMCLen{}}}{\bf \#API}  
           & \multicolumn{2}{>{\centering\arraybackslash}p{\TabMCLen{}}}{\bf \#Partial Op.} 
           & \multicolumn{2}{>{\centering\arraybackslash}p{2.5cm}}{\bf \#Record}  \\
           \cmidrule(lr){2-3} \cmidrule(lr){4-5} \cmidrule(lr){6-7}
           & \TableTorch & \TableTF & \TableTorch & \TableTF &\TableTorch & \TableTF \\
    \hline
        Collected          & 758 & 248      &  ---   &  ---     &  63,136   &   33,973  \\
        Filtering        & \multirow{2}*{681} & \multirow{2}*{214}  &  \multirow{2}*{5,875} & \multirow{2}*{1,799} &  29,589   &   12,908  \\
        Augment.         &                    &                     &                       &     &  1,041,459  &  303,314  \\
    \hline
        Inference        &    620 & 192     &  4,475 & 1,507     &  ---  &  --- \\
        Fuzzing$^\top$   &    604 & 185     &  4,186 & 1,434     &  ---  &  --- \\
        Fuzzing$^\bot$   &    582 & 176     &  4,144 & 1,415     &  ---  &  --- \\
    \hline
    \end{tabular}
\end{table}

\begin{figure}
    \centering
    \includegraphics[width=\linewidth]{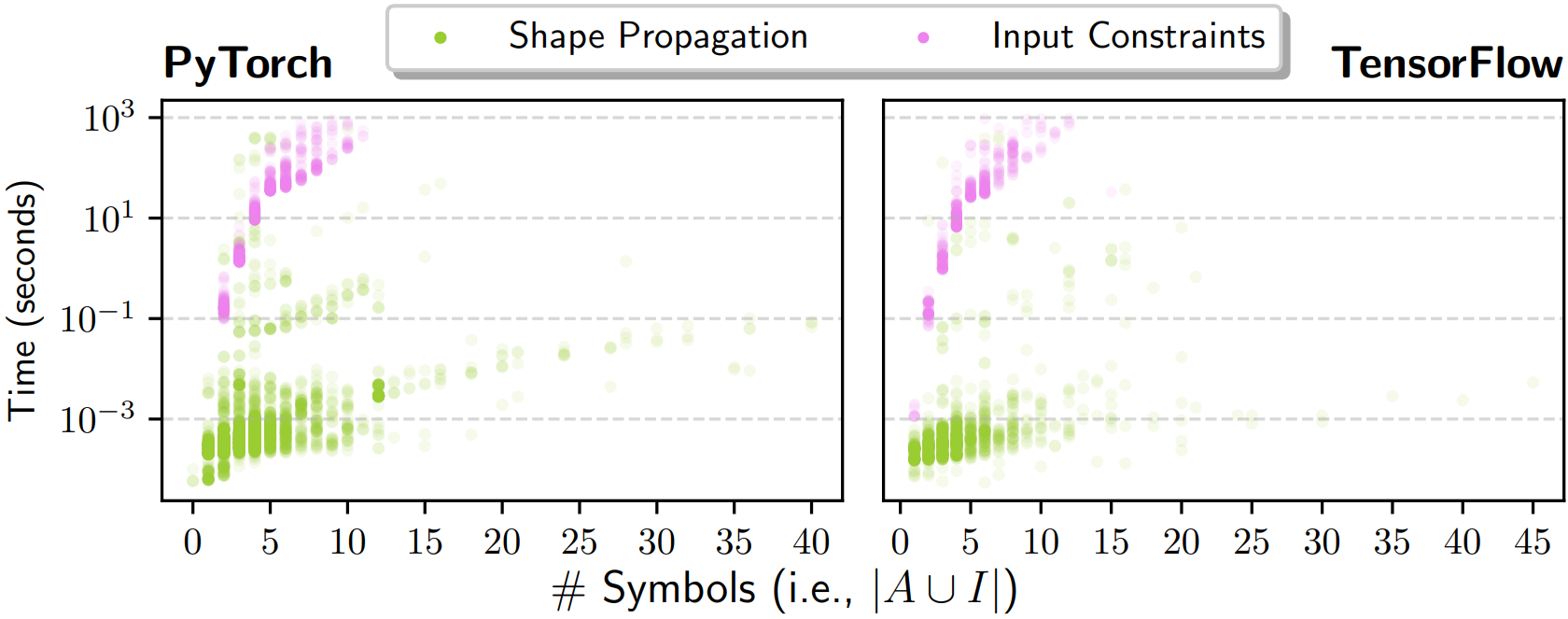}
    \caption{Inference time and \#symbols of inferred rules\protect\footnotemark.}
    \label{fig:scalability}
\end{figure}

\footnotetext{The data points of input constraints are fewer because some \ruleInst{} have an empty set of input constraints and are not visualized.}

\begin{table}
\caption{Number of inferred shape propagation rules in 1000s.}\label{tab:rosette}
    \centering
    \begin{tabular}{c 
    >{\raggedleft\arraybackslash}p{0.7cm}
    >{\raggedleft\arraybackslash}p{0.7cm}
    >{\raggedleft\arraybackslash}p{0.7cm}
    >{\raggedleft\arraybackslash}p{0.7cm}
    >{\raggedleft\arraybackslash}p{0.9cm}
    >{\raggedleft\arraybackslash}p{0.9cm}
    }
    \hline
         & \multicolumn{4}{c}{Inferred}   & \multirow{2}*{Timeout}  & \multirow{2}*{Unsat.}  \\
         \cmidrule(lr){2-5}
         & <1s & <10s     & <100s    & <1000s      &          &         \\
    \hline
 \sys     & 4,660 & 4,700        &  4,716       &   4,758         &  994       &  123       \\
 \Rosette & 0     & 83           &  2,832       &   4,461         &  1,414     &  0         \\
    \hline
    \end{tabular}
\end{table}

\noindent\textbf{Scalability.}
Figure~\ref{fig:scalability} depicts the distribution of inferred rules in terms of the inference time and symbol size.
 It shows 95\% of inferred \ruleInsts{} have less than 10 (PT) / 11 (TF) symbols, indicating the problem size is small overall.
Additionally, inferring shape propagation rules is highly affordable (\ie 95\% of them can be solved within 17ms).
However, input constraint inference tends to be more costly, since predicate candidates are thoroughly enumerated in Algorithm~\ref{algo:icons} while shape propagation terminates when the first feasible solution is found.
Meanwhile, the inference time of input constraints grows with \# symbols because \ruleInsts{} with more symbols incur more expressions to validate (Table~\ref{tab:prune}).

Table~\ref{tab:rosette} compares \sys and \Rosette in inferring shape propagation for \ruleInst{} in PyTorch.
It shows that 79\% of \ruleInsts{} are inferred by \sys within one second,
while \Rosette takes over 10 seconds for 99\% of cases.
Specifically, the ``unsat'' presents the number of cases where no expression is applicable after the \emph{full} enumeration.
\Rosette has zero ``unsat'' rules because it fails to finish the enumeration within 1000s.

\noindent\textbf{Impact of pruning.}
Table~\ref{tab:prune} shows the number of expressions after being pruned by different methods.
For example, for an \ruleInst{} with six symbols (\ie $|\attr\cup\dinp|=6$), without any pruning there are hundreds of billions of expressions to verify (\ie $1.86\times 10^{11}$, recall that we have at most 5 operations).
Our pruning method (\ie the ``both'') is able to prune the search space smaller by $478\times$.
The ablation study shows that both \emph{equivalence} and \emph{rarity} can sufficiently contribute to shrinking the search space even if applied independently.
Interestingly, we found that our method prunes even better pruning ratio for fewer symbols,
\eg $1.64\times10^{-5}$\% \emph{un}pruned for one symbol and 0.21\% \emph{un}pruned for six symbols.
This is a good news since from Figure~\ref{fig:scalability} we see the number of symbols tend to be small, \ie approximately clustered within 1 to 10.

\begin{table}[]
    \centering
    \caption{Sizes of expressions after pruning.}\label{tab:prune}
    \begin{tabular}{c |
    >{\raggedright\arraybackslash}p{1.4cm} 
    >{\raggedright\arraybackslash}p{1.4cm} 
    >{\raggedright\arraybackslash}p{1.4cm}
    >{\raggedright\arraybackslash}p{1.4cm}
    }
    \hline
                  $|\attr\cup\dinp|$ &  None   & Equiv. & Rarity & Both \\
    \hline
                      1 & $4.77\times 10^8$ %
                        & $5.35\times 10^6$ %
                        & $3.54\times 10^5$ %
                        & 78
                        \\
                      2 & $2.88\times 10^9$ 
                        & $9.04\times 10^7$ 
                        & $5.90\times 10^6$
                        & $1.25\times 10^5$ \\
                      3 & $1.11\times 10^{10}$ 
                        & $5.18\times 10^8$
                        & $5.22\times 10^7$ 
                        & $2.28\times 10^6$ \\
                      4 & $3.32\times 10^{10}$ 
                        & $1.91\times 10^9$
                        & $3.13\times 10^8$  
                        & $1.94\times 10^7$ \\
                      5 & $8.36\times 10^{10}$
                        & $5.49\times 10^9$
                        & $1.40\times 10^9$  
                        & $1.06\times 10^8$ \\
         \hline
                      6 & $1.86\times 10^{11}$
                        & $1.33\times 10^{10}$
                        & $4.66\times 10^9$  
                        & $3.89\times 10^8$ \\
         Percent.  & 100\%            & 7.15\%       &  2.51\%              & 0.21\%    \\
    \hline
    \end{tabular}
\end{table}

\noindent\textbf{Examples.}
We found most \ruleInsts{} have very simple rules being inferred, \eg 70\% (PT) / 51\% (TF) of which only have expressions with at most one symbol.
Nevertheless, we show some complicated representatives to indicate the capability of \sys's rule synthesizer.
Given an arbitrary input tensor \texttt{x}, say its shape is $[i_1, i_2, i_3, i_4]$, both \texttt{x.flatten()} and \texttt{torch.ravel(x)} flatten the tensor into a 1-D array whose shape is $[i_1 \times i_2 \times i_3 \times i_4]$, which can be inferred by us.
For more complicated cases such as \texttt{x.unfold(dim, size, step)}, the shape propagation of $o_{\texttt{dim}} = 1+\frac{(i_{\texttt{dim}}-\texttt{size})}{\texttt{step}}$ can also be correctly learnt\footnote{Division demonstrated in this paragraph is floor division, \ie for integers.}.
There are also cases where partially correct rules are learnt.
Consider \texttt{avg\_pool3d(x, ksize=[kT, kH, kW], pad=[pT, pH, pW])} where both the input and output have five dimensions (\ie $[i_N,i_C,i_T,i_H,i_W]\to[o_N,o_C,o_T,o_H,o_W]$).
While \sys correctly inferred that $o_T = \frac{i_T + 2\texttt{pT}}{\texttt{kT}}$, it overfits the H and W dimension with $o_H = \frac{i_H}{\texttt{kH}} + \min(1, \texttt{pH})$, due to the lack of records where $\frac{i_H}{\texttt{kH}} + \min(1, \texttt{pH}) \neq \frac{i_H + 2\texttt{pH}}{\texttt{kH}}$.
However, in our 4-hour fuzzing the corresponding \ruleInsts{} were used and did not lead to any invalid models.
This is because this incorrect expression still gets a good chance of being valid.
For example, \nnsmith{}~\cite{nnsmith} reveals that SMT solvers like Z3~\cite{moura2008z3} tend to return boundary models, \eg \texttt{pH} = 0 which makes the two equations equivalent.
There are of course uninferred operator rules. %
For example, one \ruleInst{} of \texttt{torch.stack} takes hundreds of input tensors, where \sys cannot handle hundreds of oriented symbols in 1000s.

\begin{table}[]
    \caption{Overview of reported bugs in four months.}\label{tab:bugs}
    \centering
    \begin{tabular}{c l | c c c}
    \hline
                & Symptom (\S\ref{sec:oracle}) & Total & Confirmed (Fixed) & Won't Fix  \\
    \hline
         \multirow{3}*{PT} & Inconsistency   & \PTBugIcons & \PTBugConfirmIcons\xspace(\PTBugFixIcons) & \PTBugWontIcons \\
                           & Runtime error  & \PTBugCE   & \PTBugConfirmCE\xspace(\PTBugFixCE) & \PTBugWontCE \\
                           & Sanitizer error & \PTBugSan   & \PTBugConfirmSan\xspace(\PTBugFixSan) & \PTBugWontSan \\
    \hline
         \multirow{3}*{TF} & Inconsistency   & \TFBugIcons & \TFBugConfirmIcons\xspace(\TFBugFixIcons) & \TFBugWontIcons \\
                           & Runtime error  & \TFBugCE   & \TFBugConfirmCE\xspace(\TFBugFixCE) & \TFBugWontCE \\
                           & Sanitizer error & \TFBugSan   & \TFBugConfirmSan\xspace(\TFBugFixSan) & \TFBugWontSan \\
    \hline
         \multicolumn{2}{c|}{Total}  & \ALLBug & \ALLBugConfirm\xspace(\ALLBugFix) & \ALLBugWont \\
    \hline
    \end{tabular}
\end{table}

\subsection{RQ3: Bug Finding}\label{sec:rq3}

\noindent\textbf{Overview and impact.}
In four months, \sys has found \textbf{\ALLBug{}} \emph{new} bugs, with \ALLBugFix{} fixed and \ALLBugConfirm{} confirmed (Table~\ref{tab:bugs}).
Links to all bug reports in this work are included in our artifact\footnote{\url{https://github.com/ise-uiuc/neuri-artifact/blob/main/docs/rq3-bug-reports.md}}.
Of these, \ALLBugFuzzing{} are found during fuzzing and \ALLBugBypro{} are byproducts (\eg crashes by counter examples in argumentation).
Among the \PTBug{} PyTorch bugs, 8 have been labelled with \textit{high priority}, constituting \textbf{10\%} (8 / 83) of all high-priority bugs for the entire PyTorch issue tracker in four months.
Besides, one \ePT{} bug has been tagged with a CVE number when there was only one other CVE published in \ePT{}.
PyTorch developers say:

\begin{shadequote}
\small 
...the bugs you've reported are \textit{high quality}, and ... don't look like specially fuzzed set that's impossible to see in practice.
They \textit{did} reveal a few common themes that are easy to encounter in practice...
\end{shadequote}

Meanwhile, to report bugs responsibly~\cite{responsible}, we discontinued bug reporting to TensorFlow when none of our first \TFBug{} reports (\TFBugConfirm{} confirmed) were fixed in a month.
Hence, the ``\TFBug{}'' bugs should be regarded as a lower bound for bug finding efficacy in TensorFlow.

\noindent\textbf{Unique bugs.}
We illustrate the patterns of exclusive PyTorch bugs (since we discontinued TensorFlow bug finding) found by \sys during fuzzing (\ie byproducts not included).
Among these \PTBugFuzzing{} fuzzing bugs in \ePT{}, 39 bugs (\ppercent{39}{\PTBugFuzzing{}}) are only manifested by models with multiple operators -- these are not able to be detected by prior single API fuzzers~\cite{wei2022free,deeprel,xie2022docter}.
Meanwhile, 41 (\ppercent{41}{\PTBugFuzzing{}}) of the bugs would not be covered by \nnsmith{} for its limited API supports\footnote{
\muffin{} shares similar limitations as \nnsmith{} in terms of limited API supports, and is not directly comparable here because it only supports TensorFlow.}.
For example, \texttt{torch.reciprocal(torch.dstack([1, 1, 1]))} ``concatenates'' the only input and gets the reciprocal, which should have returned \texttt{[1, 1, 1]}.
However, after compilation the third output element becomes non-deterministic.
This is confirmed to be a miscompilation bug\footnote{\url{https://github.com/pytorch/pytorch/issues/93078}} (now fixed) where the C kernel function generated by PyTorch has erroneous pointer aliases for input and output buffers.
This inconsistency bug is neither detectable by single-API testers nor \nnsmith{} (unsupported APIs). 

In addition, 17 bugs (\ppercent{17}{\PTBugFuzzing{}}) are exclusive to \sysi, \ie the shapes and attributes of the bug-inducing models are not directly obtained from the records, but by solving constraints from inferred rules.
It shows that enabling rule inference, though not bringing surprising coverage improvement (\S\ref{sec:rq1}), does help find more bugs.
For instance, a \emph{high-priority} bug detected by \sys\footnote{\url{https://github.com/pytorch/pytorch/issues/93274}} (now fixed) requires the input shape of \texttt{torch.histogramdd} to be specifically \texttt{[5, 6]} for triggering a compiler failure.
The input shape, \ie \texttt{[5, 6]}, comes from the solver-provided model and none of the six collected records of \texttt{torch.histogramdd} can trigger the bug.
As another example, the \texttt{*-unfold-abs\_} model pattern (``\texttt{*}'' means any operators and the ``\texttt{\_}'' in \texttt{abs\_} means it is an in-place operation) can manifest a result inconsistency bug\footnote{\url{https://github.com/pytorch/pytorch/issues/98143}} (now fixed) since the graph functionalization in the \ePT{} compiler was not able to identify certain operator patterns that have memory overlapping.
Specifically, it is detected by using a set of solver-provided arguments, \ie \texttt{tensor.unfold(1, 3, 2)}.
Notably, both examples here require operators that are unavailable in \nnsmith{}, showing that operator diversity further powers model diversity to detect more bugs.

Furthermore, one heap-overflow bug is assigned by PyTorch with a CVE identification number (\texttt{GHSA-6655-44g2-4gc8}) due to its security impact.
This bug is induced by \texttt{searchsorted(arr, val, sorted\_idx)}, which aims to binary search \texttt{val} in \texttt{arr} (\eg an array) where \texttt{sorted\_idx} are the sorted indices of \texttt{val}.
Specifically, boundary checks for \texttt{sorted\_idx} were absent in the previous implementation. 
Therefore, a large enough index, if ``lucky'', can lead to a segmentation fault, terminating the program without further impact.
However, a carefully designed index allows attackers to access and steal data from other memory addresses besides the array range when performing the binary search.

\noindent\textbf{``Won't-fix'' bugs.}
Three of our reports are rejected or deprioritized.
For example, both an inconsistency bug in \texttt{tf.cast}\footnote{\url{https://www.tensorflow.org/api_docs/python/tf/cast}}
and
a crash bug in \texttt{torch.linalg.eigvals}~\cite{linalgacc}
were rejected for using NaNs as input, incurring undefined behaviours.
Another PyTorch JIT bug was deprioritized because developers suggested us to use and test the new compiler~\cite{dynamo} (and consequently we did).

\section{Related Work}

In recent years, fuzzing~\cite{miller1990empirical} has been extensively studied for testing DL libraries and DL compilers, which can be mainly categorized into operator and model levels.
Operator-level techniques~\cite{wei2022free,deeprel,xie2022docter} aim to test each DL API in isolation. 
\freefuzz~\cite{wei2022free}, a fully automated operator-level fuzzer for testing DL libraries (such as TensorFlow and PyTorch), collects DL API traces from sources such as developer tests and model zoo, and further mutates the traced inputs to generate additional valid/invalid test-cases for fuzzing each operator.
Similarly, \doctor~\cite{xie2022docter} also aims to test each DL operator individually, by extracting their input constraints from documentation, incurring manual inspection of the mined rules for 30\% of the API arguments.
Nonetheless, the input constraints for each argument are defined by \doctor as a potential set of types and values, which cannot model fine-grained shape constraints.
Additionally, recent work has also been improving the oracles of DL system testing via API relation~\cite{deeprel} and gradient checking~\cite{nablafuzz}.
Though effective in bug finding, operator-level fuzzing hardly uncover bugs induced by multiple operators together, \eg bugs in DL compilers.

Model-level fuzzing techniques generate DL models with multiple operators. %
The pioneer \cradle~\cite{cradle} directly runs pre-built DL models programmed in Keras~\cite{keras} and cross-check results from various backends.
Built on \cradle, \lemon~\cite{lemon} and \audee~\cite{guo2020audee} generate models via pre-defined mutation rules.
Furthermore, \muffin~\cite{muffin} performs layer-by-layer model generation for testing both training and inference.
Recently, \nnsmith~\cite{nnsmith} annotates each operator with input constraints and shape transformation, and generates valid models aided by SMT solving.
While they complement operator-level fuzzing,
the model mutation/generation rules are restrictive, \eg they typically only target naive shape-preserving operators~\cite{lemon,muffin,luo2021graph}, or require certain manual annotations~\cite{nnsmith}, leading to a limited set of operators being used. %
This work proposes to infer such operator rules, and then apply them to generate valid models with all possible operators.
Our work can cover as many operators as operator-level fuzzing while generating valid models with covered operators \emph{fully automatically}, \ie a step forward for bridging the gap between operator- and model-level fuzzing for DL systems. 

More recently, there has been concurrent work~\cite{titanfuzz, fuzzgpt} on directly leveraging large language models (LLMs) to synthesize Python programs to construct valid DL models. 
Such techniques aim to \emph{implicitly} solve the validity constraints via directly learning from valid samples.
Compared with LLM-based model generation, despite the technical complexity, by \emph{explicitly} solving the constraints,
\sys provides stronger validity guarantee within the covered model space and is more affordable (\ie <100ms per model on CPU).
Meanwhile, LLMs, trained over billions of lines of codes, can be used to easily test beyond the model space carefully crafted and exhaustively explored by \sys.
Therefore, these two approaches can be further combined for maximized fuzzing in the future.

Lastly, program synthesis has been used by related areas such as synthesizing user-facing tensor-manipulation programs~\cite{shi2022tf, ni2021soar, zhou2022intent}. 
Similar to \nnsmith, they also require manual operator specifications.
This paper applies inductive program synthesis~\cite{Winston:1970,Lau:1998} to infer such specifications, and can potentially improve all methods targeting model generation, \eg DL system fuzzing, tensor-manipulation program synthesis and neural architecture search~\cite{elsken2019neural}.

\section{Conclusion}
In this paper, we present \sys,
the first approach to automatically infer operator rules for diversifying model generation in order to test DL systems.
\sys generates test-cases from structurally valid models composed by diverse operators for exercising deeper system behaviours.
The primary source of the diversity comes from our automated rule inference engine and concolic model generator.
The rule inference engine inductively and efficiently discovers operator rules for generating valid models symbolically.
Meanwhile, our concolic model generator can further make use of concrete operator invocations in combination with the symbolic operators to maximize the model diversity. 
As a result, \sys finds many \emph{high-priority} and -\emph{quality} bugs appreciated by DL-framework developers.
Additionally, \sys is practical and promising for long-term fuzzing -- high-quality test-cases can be generated in milliseconds on a single CPU thread and new operators can always be automatically integrated.
To date, \sys has already detected \ALLBug{} \emph{new} bugs for PyTorch and TensorFlow, with \ALLBugConfirm{} fixed or confirmed.

\section*{Data Availability}

The artifact is available at \url{https://github.com/ise-uiuc/neuri-artifact}.

\section*{Acknowledgments}
This work was partially supported by NSF grants CCF-2131943 and CCF-2141474, as well as research awards from Google and Meta.
We thank Jun Yang for providing valuable proofreading assistance.

\bibliographystyle{ACM-Reference-Format}
\bibliography{reference}

\end{document}